\documentclass[twocolumn,showpacs,preprintnumbers,amsmath,amssymb]{revtex4}

\usepackage{graphicx}
\usepackage{dcolumn}
\usepackage{bm}

\usepackage{epsfig}
\begin{document}

\title{\bf Relativistic Structure, Stability
and Gravitational Collapse of 
Charged  Neutron Stars}

\author{Cristian R. Ghezzi\thanks{Electronic mail: ghezzi@ime.unicamp.br}}

\address{\it Department of Applied Mathematics, \\
Instituto de Matem\'atica, Estat\'{\i}stica e Computa\c c\~ao Cient\'{\i}fica,\\ 
Universidade Estadual de Campinas,\\ Campinas, S\~ao Paulo, Brazil}

\date{\today}
\begin{abstract}
Charged stars have the potential of becoming charged 
black holes or even naked singularities.
It is presented a set of numerical solutions of the 
Tolman-Oppenheimer-Volkov
equations that represents spherical charged compact 
 stars
in hydrostatic equilibrium. The stellar models obtained are
evolved forward in time integrating  the 
Einstein-Maxwell field equations.
It is assumed an equation of state of a neutron gas at zero temperature. 
The charge distribution is taken as been
proportional to the rest mass density distribution. 
The set of solutions
present an unstable branch, even with charge to mass ratios arbitrarily close to the extremum case. 
It is performed a direct check of the stability of the solutions  under strong perturbations, and for different
 values of the charge to mass ratio.
The stars that are in the stable
branch oscillates and do not collapse, while models 
in the unstable branch collapse 
directly to form black holes. Stars with a charge greater or equal than the extreme value explode.
  When a charged star is suddenly discharged, it don't necessarily collapse to form a  black hole.
A non-linear effect that gives rise to the formation of an external shell of matter (see Ghezzi and Letelier 2005), is negligible in the present simulations.
The results are in agreement with the third law of black hole thermodynamics and with the cosmic censorship conjecture.
 
\end{abstract}
\maketitle

\section{INTRODUCTION}
The study of charged relativistic 
fluid balls attract the interest
of researchers of different areas of physics and astrophysics. 
There  exists a 
general consensus that  
astrophysical objects with large amounts of charge can not exist in nature \cite{eddington}, \cite{glendenning}.
This point of view had been challenged by several researchers \cite{bally}, \cite{olson1}, \cite{olson2}, \cite{mosquera}. 
It cannot be discarded  the possibility
that during the gravitational collapse or during an accretion process onto a compact object the matter acquires  large amounts of electric charge. This has been considered in \cite{diego} and \cite{shvartsman}. 

In this paper we study the stellar structure and temporal evolution of compact charged fluid spheres, independently on the mechanism by which the matter acquires an electric charge.

From a pure theoretical point of view, the collapse of charged fluid balls and shells are  connected with the laws
of black hole thermodynamics.  
The third law of black hole thermodynamics states that no  process can reduce the surface gravity of a black hole to zero in a finite advanced time \cite{bardeen}, \cite{poisson}, \cite{novikov}. 
The surface gravity, $\kappa$, plays the role of a temperature while the area of the event horizon is equivalent to the entropy. Israel \cite{israel} gave a formulation and
proof of the third law .
It was demonstrated that the
laws of black hole thermodynamics are analogous to the
common  laws of  thermodynamics \cite{novikov}.
Translated to the physics of fluid collapse,
the third law implies the impossibility of forming an extremal black hole, for which $\kappa=0$. An 
extremal black hole has a total charge $Q$ : $Q=\sqrt{G}M$,  
where ${G}$ is the gravitational constant, and $M$ is the total mass of the black hole. The extremal black hole has not an horizon and constitutes a naked singularity, contradicting the cosmic censorship hypothesis.
So, in this paper we explore the possibility of forming an extremal black hole from the collapse of a (charged) compact object. We will show that the answer is no.

The relativistic equations for the collapse of a charged
fluid ball were obtained by
Bekenstein \cite{bekenstein}.

So far
in the literature,  as long as the author knows, it was studied 
the dynamics of charged shells or scalar fields
 falling onto an already formed charged black hole (see \cite{boulware},  \cite{chase}, \cite{piran}, and references therein), and solutions of the Tolman-Oppenheimer-Volkoff \cite{oppen} equations for charged stars (see \cite{anninos}, \cite{defelice}, \cite{ray}, \cite{zhang},  etc). 
 However, in the stellar collapse, the collapsing matter  determines the background geometry and the metric of space-time is changing with time as the star collapses.
Moreover, in general relativity, 
the pressure of the fluid contributes to
 the gravitational field and strengthen it.
From the present study it is possible to understand if the Coulomb repulsion will prevent or not the total collapse of a charged fluid ball, and which are the stability
limits for a charged relativistic star. We emphasize that we will not be concerned here with the mechanism of electric charge generation, nor with effects that could neutralize or discharge the star.

In the present paper,  
the equations for the evolution of charged fluid spheres
in Schwarzchild-like coordinates are presented in Sec. \ref{relatequ}. In the Sec. \ref{sec2}, we re-derived the equations of hydrostatic equilibrium and  the relativistic equations for the temporal 
evolution  of charged spheres  in a form 
closer to that obtained by May and White 
\cite{may}, \cite{may2} (for the case of zero charge).
The equations obtained are  
well suited for performing a finite
difference scheme, and allows a direct implementation of very well known numerical
techniques \cite{ghezzi}, \cite{may}, \cite{may2}. 
The formalism is also compatible
with the Bekenstein equations \cite{bekenstein}, and with
the Misner and Sharp equations \cite{misner}  for the case of zero charge.

The calculations performed in this study are rather lengthy in general and 
we tried to be the more self-consistent as possible. However, we apologize that in most of the paper we only give hints for the computation of the intermediate algebraic steps in order to keep the paper at a reasonable length.

The matching conditions between the exterior and interior
solutions of the Einstein-Maxwell equations are given in the 
Sec. \ref{match}.

For completeness, in Sec. \ref{seceos} it is considered an equation of state of a zero temperature
 neutron gas. 

Two codes were built and are used in the present study: one to obtain neutron stars models in hydrostatic equilibrium, presented in the Sec. \ref{sectov};
and the other code is to evolve the stars forward in time, integrating the Einstein-Maxwell equations. This code is introduced in Sec. \ref{sechydro}. 
In all the simulations, the charge distribution is taken proportional to the rest mass distribution.

We found that although the total binding energy grows with the total amount of charge, the binding energy per nucleon tends to zero as the charge tends to the extremal value.  This is related to the impossibility of obtaining bounded solutions of extremely charged fluid configurations. In addition, it results that the upper limit for the total amount of charge could be slightly lower than the extremal value.
This is discussed in the Sec. \ref{secresults}.

In the Sec. \ref{secresults}, is  discussed the stability and evolution of the stellar models.  We study, as well, the evolution of stars that were  suddenly discharged to take into account the case of charged spheres being a metastable state of more complex scenarios. 

In a recent work Ghezzi and Letelier \cite{ghezzi} presented a numerical study of the collapse of charged fluid spheres  with a polytropic equation of state, 
and with an initial uniform distribution of energy density and charge.
They found that during the evolution a shell of higher density is formed 
near the surface of the imploding star.  
From that work we can not say if a shell will always form  when changing
 the initial conditions and the scenario for the collapse.
In the present study, we checked that the effect is negligible during the collapse of a charged 
neutron star (see Sec. \ref{secresults}).

In the Sec. \ref{secremarks}, we end  with some final remarks.

\section{Relativistic equations\label{relatequ}}
Considering  a spherical symmetric fluid ball, the   line element in Schwarzchild-like, or standard form, is 
(see \cite{weinberg}, \cite{landau}, \cite{bekenstein}):
\begin{equation}
\label{metrica0}
ds^2=-e^{\nu} dt^2+e^{\lambda} dr^2 
+r^2 d\theta^2 + r^2 {\rm sin}^2 \theta d\phi^2
\end{equation}
An observer moving  radially have a 4-velocity  \mbox{$u^{\nu}=(u^0,u^1,0,0)$}, and the electric 4-current  is \mbox{$j^{\nu}=(j^0,j^1,0,0)$}. 
The energy-momentum tensor of the charged fluid is given by:
\begin{eqnarray}
&&{T}^{\nu \mu}=(\delta+P) u^{\mu} u^{\nu}+P\,g^{\mu \nu}+\nonumber \\
&&\hspace{1cm}
\frac{1}{4 \pi} [ F^{\mu \alpha} F^{\nu}_{\alpha}-
\frac{1}{4} g^{\mu \nu} 
F^{\alpha \beta} F_{\alpha \beta} ]
\end{eqnarray}
where $\delta$ is the density of mass-energy, $P$ is the scalar
pressure, and $F^{\mu \alpha}$ is the electromagnetic tensor.
The electromagnetic field satisfies the Maxwell
equations:
\begin{equation}
\label{Maxwell}
F^{\mu \nu}_{;\nu}=4 \pi j^{\mu}\,,
\end{equation}
and
\begin{equation}
F_{[\alpha \beta,\gamma]}=0\,.
\end{equation}
Only the radial component  $F^{0 1}$ is non-zero, and the last equation is satisfied if $F^{0 1}=-F^{1 0}$.
The covariant derivative in Eq. ($\ref{Maxwell}$) can be written as 
\begin{equation}
\label{Maxwell2}
\frac{1}{(-g)^{1/2}} \bigl[(-g^{1/2}) F^{\mu \nu}\bigr]_{,\nu}=4 \pi j^{\mu}\,,
\end{equation}
with $(-g)^{1/2}=r^2 e^{(\lambda+\nu)/2}$.
Defining \cite{bekenstein}
\begin{equation}
\alpha=\frac{1}{2} (\lambda+\nu)\,,
\end{equation}
the Eq. (\ref{Maxwell2}) gives \cite{bekenstein}
\begin{equation}
\label{Mw1}
\frac{d(r^2 e^{\alpha} F^{01})}{dr}=
4 \pi r^2 j^0  e^{\alpha}\,,
\end{equation}
and
\begin{equation}
\label{Mw2}
\frac{d(r^2 e^{\alpha} F^{01})}{dt}=
-4 \pi r^2 j^1  e^{\alpha}\,.
\end{equation}

Integration of Eq. (\ref{Mw1}) gives \cite{bekenstein}
\begin{equation}
\label{campoe}
F^{01}=e^{-\alpha} Q(t,r)/r^2\,,
\end{equation}
where
\begin{equation}
Q(t,r)=\int^{r}_{0}  4 \pi r^2 j^0 e^{\alpha} dr\,,
\end{equation}
and from Eq. (\ref{Mw2})
\begin{equation}
\label{dqdt}
\frac{dQ}{dt}=
-4 \pi r^2 j^1  e^{\alpha} \,.
\end{equation}
From this equation we see that the charge is a constant $e$
outside the fluid ball, where $j^1=0$.
 Far away from the sphere, where $e^{\alpha} \rightarrow 1$ (see below),
 the Eq. (\ref{campoe}) gives the electric field of a classical charged particle. 
 
The Einstein's equations to be solved are:
\begin{equation}
R^{\mu \nu}-\frac{1}{2} g^{\mu \nu} R= \frac{8 \pi G}{c^4} T^{\mu \nu} \,,
\end{equation}
here (and only here), 
$R$ denotes the scalar curvature, and $R^{\mu \nu}$
is the Ricci tensor. 

The  components of the Einstein equations are (see  \cite{bekenstein}, \cite{landau}):
\begin{eqnarray}
\label{E1}
&&T^{0}_{0}\,\,\mbox{:}\,\,\frac{Q^2}{r^4}-8 \pi\, [(\delta+P)\,u^0u_0+P]=  \nonumber\\
&&\hspace{0.8cm} e^{-\lambda} \,\biggl(\frac{\lambda'}{r}-\frac{1}{r^2}\biggr)+\frac{1}{r^2}\,,
\end{eqnarray}

\begin{eqnarray}
\label{E2}
&&T^{1}_{1}\,\,\mbox{:}\,\,\frac{Q^2}{r^4}-8 \pi\, [(\delta+P)\,u^1 u_1+P]= \nonumber\\
&&\hspace{0.8cm} -\,e^{-\lambda}\, \biggl(\frac{\nu'}{r}+\frac{1}{r^2}\biggr)+\frac{1}{r^2}\,,
\end{eqnarray}

\begin{equation}
\label{E3}
T^{1}_{0}\,\,\mbox{:}\,\,8 \pi\, (\delta+P)\,u^1 u_0=
e^{-\lambda} \dot{\lambda}/r\,.
\end{equation}

Multiplying the right hand side of Eq. (\ref{E1}) by
$r^2$ and rearranging, the equation can be 
cast in the form
\begin{equation}
\label{RE1}
-\frac{d(r e^{-\lambda})}{dr}+1=\frac{Q^2}{r^2}
-8 \pi\, [(\delta+P)\,u^0 u_0+P]\, r^2\,.
\end{equation}
Setting $e^{-\lambda}=1+f+g$, with  $f=-2m/r$ and $g=Q^2/r^2$:  
\begin{equation}
\label{gamma0}
e^{-\lambda}=1-\frac{2m}{r}+\frac{Q^2}{r^2}\,,
\end{equation}
the Eq. (\ref{RE1}) becomes
\begin{equation}
\label{M1}
\frac{dm}{dr}-\frac{1}{2} \frac{d(Q^2/r)}{dr}=\frac{1}{2} \frac{Q^2}{r^2} 
-4 \pi \, [(\delta+P)\, u^0 u_0 +P]\, r^2\,.
\end{equation}
Integrating this equation, we get the equation for 
the mass \cite{bekenstein}:
\begin{eqnarray}
\label{M1}
&&m=-\frac{4 \pi}{c^2} \int^r_0 [(\delta+P)\, u^0 u_0 +P]\, r^2\, dr \nonumber\\
&&\hspace{0.8cm}+\frac{1}{2 c^2} \frac{Q^2}{r}+
\frac{1}{2 c^2} \int^r_0 \frac{Q^2}{r^2} dr+m_0\,, 
\end{eqnarray}
where $m_0$ is an integration constant, and can be taken as
zero for the purpose of this work.

 Outside the fluid ball, the mass  do not depend on $r$ and then: $m:m(t,r_s)$ for $r>r_s$, where $r_s$ is the coordinate $r$ of the surface of the sphere.
Actually the mass do not depend on $t$  for $r>r_s$; 
from Eqs. (\ref{E3}) and (\ref{gamma0}): 
\begin{equation}
\label{dmdt}
\frac{dm}{dt}=\frac{Q}{r}\frac{dQ}{dt}+4 \pi\, (\delta+P)\, r^2\, u_0 u^1\,,
\end{equation}
as  $Q$ is constant and $\delta=P=0$ outside the fluid ball, 
 $m$ is independent of $t$.

Subtracting Eq. (\ref{E1}) from Eq. (\ref{E2}) it is
\begin{equation}
\label{metrica0b}
\frac{e^{-\lambda}}{r} \frac{d(\lambda+\nu)}{dr}=
8 \pi\, (\delta+P) \,(u^1 u_1-u^0 u_0)\,.
\end{equation}
So, $\lambda+\nu$ is
independent of $r$. To get an asymptotic flat
solution it must be $\lambda+\nu \rightarrow 0$ as $r \rightarrow \infty$.
Thus a solution of Eq. (\ref{metrica0b}) is \cite{bekenstein}
\begin{equation}
\label{coefmetrica1}
\lambda=-\nu\,\,\,\,\,\,\,\,{\rm for}\,\,\,\,r>r_s 
\end{equation}
 
 Using
the Eq. (\ref{gamma0}), (\ref{coefmetrica1}), and substituting 
into Eq. (\ref{metrica0}), the line 
element outside the ball is
\begin{eqnarray}
\label{metrica1}
&ds^2=-\bigl(1-\frac{2M}{r}+\frac{Q^2}{r^2}\bigr)\, dt^2
+ \bigl(1-\frac{2M}{r}+\frac{Q^2}{r^2}\bigr)^{-1} \,dr^2
 \nonumber\\
&+r^2 d\theta^2 + r^2\,{\rm sin}^2 \theta\, d\phi^2\,,
\end{eqnarray}
which is the Reissner-Nordstr\"om space-time. The gravitational mass is $M=m(r_s)=constant$.
 We see that although the (interior) mass 
distribution and the metric
depends on time, the external space-time is static. This is the Birkhoff theorem, and implies that
a spherical distribution of mass and charge cannot emit
gravitational waves.

\section{\label{sec2}Equations in co-moving coordinates}
In this section we will specialize the equations for the case of a coordinate system co-moving with the fluid. We will use a  notation closer
to that used in the  May \& White papers  \cite{may}.

The 4-velocity for an observer co-moving with the
fluid is
$u^{\mu}=(a^{-1},0,0,0)$, and the measured 4-current is $j^{\mu}=(j^0,0,0,0)$. The 4-velocity satisfies: 
$u^{\mu} u_{\mu}=c^2$.

The  line element for co-moving coordinates in ``standard" form (see \cite{weinberg}, Pag. 336) is:
\begin{eqnarray}
\label{comline}
&&ds^2=a(t,\mu)^2 c^2 dt^2-b(t,\mu)^2 d\mu^2 \nonumber \\
&&\hspace{1cm}-R(t,\mu)^2(d\theta+{\rm sin}^2\theta d\phi^2)\,,
\end{eqnarray}
note that now $a$, $b$, and $R$ are functions of $\mu$ and $t$.
We will see in the next subsection that the coordinate $\mu$ can be chosen to be the total rest mass inside a sphere,  such  that
each spherical layer of matter can be labeled by the rest mass it contains. 

The stress tensor in co-moving coordinates is obtained using the solution
of the Maxwell equations (\ref{campoe}):
\begin{eqnarray}
&&T^0_0=\delta c^2+\frac{Q^2}{8 \pi R^4}\\
&&T^1_1=-P+\frac{Q^2}{8 \pi R^4}, \\
&&T^2_2=T^3_3=-P-\frac{Q^2}{8 \pi R^4}\\
&&T^0_1=T^1_0=0\,.
\end{eqnarray}
From now on we will restore in the equations the gravitational constant $G$ and the speed of light $c$.

 Similarly as was defined above, $\delta\,c^2$  is the density of 
energy in $[{\rm dyn/cm}^2]$ and $P$ is the scalar pressure
in the same units.
It will be useful to perform the split  \footnote{See for example, the Refs. \cite{zeldovich}, \cite{shapiro}, and \cite{may}. }
\begin{equation}
\label{split}
\delta=\rho\, (1+\epsilon/c^2)\,,
\end{equation}
 where
$\rho$ is the rest mass density and $\rho\, \epsilon$
is the internal energy of the gas.

\subsection{Equation of particle conservation}
We want to set the radial coordinate equal to the rest mass
enclosed by co-moving spherical layers.
We performed a calculation analogous to that of May and White \cite{may}, 
and used the same notation for clarity and comparison. 

In this work is assumed that there is only one species
of particles, and there are no particles created nor 
destroyed. 
The number density of particles will be
denoted by $n$ and the rest mass density by $\rho$. If the rest
mass of the particles is $m_n$ then:
\begin{equation}
\rho=n \,m_n\,.
\end{equation}
The conservation of the baryon number current is expressed :
\begin{equation}
n \,u^{\nu}_{;\nu}=0\,
\end{equation}
 or equivalently:
\begin{equation}
\label{baryons}
\rho \,u^{\nu}_{;\nu}=0\,.
\end{equation}

The line element in  spherical coordinates written
in ``standard" form 
\footnote{See Ref. \cite{landau}, Pag. 304, and Ref. \cite{weinberg}, Pag 336.} is :
\begin{eqnarray}
\label{comline2}
&&ds^2=a(t,R)^2 \,c^2\, dt^2-b(t,R)^2\, dR^2 \nonumber \\
&&\hspace{1cm}-R^2\,(d\theta+{\rm sin}^2\theta\, d\phi^2)\,,
\end{eqnarray}
observe that this line element is not identical with that given in
 Eq. (\ref{comline}).

The conservation of baryons in this frame is reduced to 
(see Eq. \ref{baryons}):
\begin{equation}
\label{dmetb}
(\rho \,R^2\,b)_{,t}=0\,,
\end{equation}
thus the quantity $(\rho \,R^2\,b)$ is a function of
$R$ only
\begin{equation}
\label{metb}
(\rho \,R^2\,b)=f(R)\,.
\end{equation}
Choosing  $f(R)=1/4\, \pi$:
\begin{equation}
\label{metricgrr}
b=1/4 \,\pi\,\rho \,R^2\,.
\end{equation}
On the other hand,  the proper mass
is 
\begin{equation}
\mu=\int_V \rho\, \sqrt{\,^{(3)} g} \,d^3x\,,
\end{equation}
where $\sqrt{\,^{(3)} g}\, d^3x=b\,R^2 \,{\rm sin} \,\theta \,dR\, d\theta \,d\phi$ is the volume element of the space section \footnote{See Ref. \cite{wald}, Pag. 126.}, and $V$ is the volume of integration.
Then
\begin{equation}
\mu=\int_0^{R_s} 4\,\pi\, \rho\, R^2\, b\, dR\,,
\end{equation}
is the proper mass enclosed by a sphere of circumference 
$2 \,\pi\, R_s$, and radial coordinate $R_s$.
With this election of coordinates: 
\begin{equation}
\label{diffs}
d\mu=dR.
\end{equation}
So we can perform the coordinate transformation 
$$(t,R,\theta,\phi) \rightarrow (t,\mu,\theta,\phi)\,,$$
and in this case the metric functions change to 
\begin{subequations}
\begin{eqnarray}
&&a(t,R) \rightarrow a'(t,\mu)\,, \\ 
&&b(t,R) \rightarrow b'(t,\mu)\,, \\
&&R \rightarrow R'(t,\mu)\,.
\end{eqnarray}
\end{subequations}
 Replacing these functions on the line element (\ref{comline2}),  
dropping the primes, and using
the Eq. (\ref{diffs}) we obtain the line element given in the Eq. (\ref{comline}). 
So, the radial Lagrangian coordinate was gauged to be the rest 
mass ``$\mu$'' of each spherical layer of matter.

\subsection{Charge conservation}
The proper time derivative of the charge is (see Eq. \ref{dqdt})
\begin{equation}
\label{chargeconsintime}
Q_{,t}=0\,,
\end{equation}
because it is possible to write
the 4-current as a product of a scalar charge density
times the 4-velocity, i.e.: $j^{\mu}=\rho_{ch}\,(u^0,0,0,0)$.
Thus the electric charge is conserved in spherical layers co-moving  with the fluid.

\subsection{Einstein-Maxwell equations in the co-moving frame}
The Einstein equations (see the Appendix)  
for the components $G_\mu^t=G_t^\mu$ are
\begin{equation}
\label{r01}
\frac{a_{,\mu}}{a} R_{,t}+ \frac{b_{,t}}{b} R_{,\mu}-
R_{,\mu t}=0\,,
\end{equation}
 where we use the notation $R_{,t}=\partial R/\partial t$, 
$R_{,\mu t}=\partial^2 R/\partial t\partial \mu$, etc.

The equation for the component $G_t^t$  is\footnote{
There are some typos in the equations of the Bekenstein's paper \cite{bekenstein}, \cite{bekenstein2}. On the Eq. 
(34) and (35) of that paper, 
the minus sign of the term $Q^2/r$, must be a plus. 
This is immediately evident 
when obtaining Eq. (42) from Eq. (35), in that work.}:
\begin{eqnarray}
\label{E00}
&&4 \pi G\,\delta\, R^2 R_{,\mu}=\frac{c^2}{2} \biggl[
R+\frac{R\,R_{,t}\,^2}{a^2\,c^2}-\frac{R\,R_{,\mu}\,^2}{b^2}
+\frac{G\, Q^2}{c^4\,R}\biggr]_{,\mu}\,  \nonumber\\
&&\hspace{0.45cm}-\frac{G Q Q_{,\mu}}{c^2 R}=G m_{,\mu}-\frac{G Q Q_{,\mu}}{c^2 R}\,,  
\end{eqnarray}
and the equation for the component $G_\mu^\mu$ is\footnote{To derive this equation, 
use the Eq. (\ref{r01}) to eliminate $a_{\mu}/a$ from the component $G_\mu^\mu$ of the 
Einstein's equations, given in the Appendix. In addition, use that $Q_{,t}=0$.}:
\begin{eqnarray}
\label{E11}
&&\frac{4 \pi G}{c^2} P R^2 R_{,t}=-\frac{c^2}{2} \biggl[
R+\frac{R\,R_{,t}^2}{a^2\;c^2}-\frac{R\,R_{,\mu}^2}{b^2}
+\frac{G\, Q^2}{c^4\,R}\biggr]_t\,\nonumber\\
&&\hspace{2.02cm}=-G m_{,t}\,. 
\end{eqnarray}

We introduced into Eqs. (\ref{E00}) and (\ref{E11}), 
the definition of the total mass:
\begin{equation}
\label{totalmass}
m(\mu,t)=4 \pi \int_0^{\mu} \delta \,R^2 R_{,\mu} d\mu 
+\frac{1}{c^2} \int_0^{\mu} \frac{Q Q_{,\mu}}{R} d\mu\,.
\end{equation}

We give further definitions that simplify the 
aspect of the equations:
\begin{eqnarray}
\label{gammadef} 
&&\Gamma=\frac{R_{,\mu}}{b}=4 \pi \rho R^2 R_{,\mu} \\
&&\hspace{0.02cm}u=\frac{R_{,t}}{a}\,. \label{udefin}
\end{eqnarray}
With this definitions an integral of 
the Eq. (\ref{E00}) and (\ref{E11}) is:
\begin{equation}
\label{gamma2}
\Gamma^2=1+\frac{u^2}{c^2}-\frac{2 m G}{R c^2}+
\frac{G Q^2}{c^4 R^2}\,,
\end{equation}
where $m$ is given by the Eq. (\ref{totalmass}).
Combining Eqs (\ref{metricgrr}) and (\ref{r01}) we get
the mass conservation equation
\begin{equation}
\label{massconservation}
\frac{(\rho R^2)_{,t}}{\rho R^2}=\frac{-a u_{,\mu}}{R_{,\mu}}\,.
\end{equation}

It can be seen that the quantity $u$ 
(see Eq. \ref{udefin}) is
the radial component of the 4-velocity of the fluid in Schwarzschild coordinates. Using the transformation rule for the contra-variant 4-vector $u^\nu$ from co-moving to Schwarzschild coordinates: ${u'}^\nu=(\partial {x'}^\nu/\partial x^\alpha)\, u^\alpha$, yields \cite{may}:
$${u'}^0=c T_{,t}/a$$
$${u'}^1=R_{,t}/a\,.$$

\subsection{Conservation of the energy and momentum}
From the Einstein equations and the contracted Bianchi identities  follows that:
\begin{equation}
T_{\nu}\,^{\mu}_{;\mu}=0\,.
\end{equation}
In  co-moving coordinates the components are:
\begin{eqnarray}
\label{EM0}
&&T_{0}\,^{\mu}_{;\mu}=\delta_{,t}+\biggl(\frac{b_{,t}}{b}
+\frac{2 R_{,t}}{R}\biggl)\biggr(\frac{P}{c^2}+\delta\biggr)=0\,, \\ 
\label{EM1}
&&T_{1}\,^{\mu}_{;\mu}=\frac{a_{,\mu}}{a}
+\frac{P_{,\mu}}{(\delta c^2+P)}- \nonumber\\
&& \hspace{0.9cm}\frac{1}{4 \pi} 
\frac{Q Q_{,\mu}}{(\delta c^2+P)\, R^4}=0\,.
\end{eqnarray}

Expanding the Eq. (\ref{dmetb}) and using the Eq. (\ref{EM0})
 we get:
\begin{equation}
\label{1stlaw}
\epsilon_{,t}=-P\, \biggl(\frac{1}{\rho}\biggr)_{,t}\,,
\end{equation}
which is identical to the non-relativistic adiabatic
energy conservation equation, and express the first law
of thermodynamics. It could be surprising for the 
reader that there is not an electromagnetic term
on this equation: this is due to the spherical
symmetry of the problem.

 The Eq.
(\ref{EM1}) can be written: 
\smallskip
\begin{eqnarray}
\label{gtt1}
&&\frac{\bigl[a (\delta+P/c^2)/\rho\bigr]_{,\mu}}{a (\delta+P/c^2)/\rho}
=\frac{1}{\bigl[(\delta+P/c^2)/\rho\bigr] c^2} \biggl[\epsilon_{,\mu} + \nonumber\\
&&\hspace{3.3cm}P \biggl(\frac{1}{\rho} \biggr)_{,\mu}
+\frac{Q Q_{,\mu}}{4 \pi R^4 \rho} \biggr]\,, 
\end{eqnarray}
\smallskip
or using the Eq. (\ref{split}) and 
the definition of the relativistic specific enthalpy (see \cite{misner}, \cite{may})
$w=(\delta+P)/\rho c^2$:
\begin{equation}
\label{enthalpy} 
w=1+\frac{\epsilon}{c^2}+\frac{P}{\rho c^2}\,,
\end{equation}
we can write the Eq. (\ref{gtt1}) in a form closer
to that used by May \& White \cite{may}:
\begin{equation}
\label{gtt2}
\frac{(a w)_{,\mu}}{a w}
=\frac{1}{w c^2} \biggl[\epsilon_{,\mu} + P \biggl(\frac{1}{\rho} \biggr)_{,\mu}
+\frac{Q Q_{,\mu}}{4 \pi R^4 \rho} \biggr]\,.
\end{equation}

\subsection{Equation of motion}
The equation of motion can be obtained expanding the
Eq. (\ref{E11}),  using the Eq. (\ref{r01}) 
to eliminate the factor
$R_{,\mu t}$, the Eq. (\ref{EM1}) to eliminate $a_{\mu}/a$ and the definitions given in the Eqs. (\ref{udefin}),
 (\ref{gammadef}), (\ref{gamma2}), (\ref{enthalpy}).
After lengthy calculations we obtain:
\begin{eqnarray}
\label{motion}
&&u_{,t}=-a \,\biggl[4 \,\pi \,R^2 \,\frac{\Gamma}{w} 
\biggl(P_{,\mu} - 
\frac{Q\,Q_{,\mu}}{4\, \pi\, R^4}\biggr)+\frac{G\,m }{R^2}+ \nonumber\\
&&\hspace{1cm}\frac{4\,\pi\, G}{c^2}\,P\, R-\frac{G\, Q^2}{c^2 \,R^3} \biggr]\,.
\end{eqnarray}
This equation reduce to the equation of motion obtained by Misner and Sharp \cite{misner} and May and White \cite{may} for the case $Q=0$, and is compatible with the equation derived by Bekenstein \cite{bekenstein}.
The Eq. (\ref{motion}) reduce to the Newtonian equation of motion of a charged fluid sphere   
letting $c \rightarrow \infty$, and consequently:
 $\Gamma \rightarrow 1$ (see Eq. \ref{gamma2}); $w\rightarrow 1$ (see Eq. \ref{enthalpy}); and $a \rightarrow 1$ (see Eq. \ref{EM1}).

\subsection{Equation of Hydrostatic Equilibrium}
The equation for charged fluid spheres in hydrostatic equilibrium 
is a generalization
of the Tolman-Oppenheimer-Volkov \cite{oppen} equation obtained  by Bekenstein \cite{bekenstein}.
We can obtain it from the equation of motion  (\ref{motion}), taking \mbox{$u=0$, and $u_{,t}=0$}.
In hydrostatic equilibrium, the factor $\Gamma$ 
(see Eq. \ref{gammadef}) is
\begin{equation}
\label{GammaHE}
\Gamma^2=1-\frac{2 m G}{R c^2}+\frac{G Q^2}{c^4 R^2}\,.
\end{equation}

Rearranging\footnote{The first two terms of 
Eq. (\ref{motion}) can be simplified using 
the chain rule:  $dP/d\mu=(dP/dR)\, R_{,\mu}$, and 
using the Eq. (\ref{gammadef}).} the Eq. (\ref{motion}) 
and using the Eqs. (\ref{enthalpy}) and (\ref{GammaHE}),
the Tolman-Oppenheimer-Volkov (TOV) equation for charged fluid spheres is:
\begin{eqnarray}
\label{TOVeq}
&&\frac{dP}{dR}=-(\delta c^2+P) \frac{\bigl(\frac{m G}{c^2}+ \frac{4\pi G}{c^{4}}  P R^3-
\frac{G Q^2}{R c^4}\bigr)}
{R \bigl(R-\frac{2 m G}{c^2}+\frac{G Q^2}{R c^4}\bigr)}+\nonumber\\
&&\hspace{0.9cm}\frac{Q}{4 \pi R^4} \frac{dQ}{dR}\,. 
\end{eqnarray}
This equation gives the well known Tolman-Oppenheimer-Volkov \cite{oppen} equation when $Q=0$.

\section{\label{match} The matching of the interior with the exterior solution}
In the Section \ref{relatequ} we find an exterior solution to the problem which agrees with the Reissner-Nordstr\"om
 solution. In the preceding section we derived an interior solution that we want to integrate with a computer. 
However,
we must show that the interior and the exterior solutions match smoothly.
The matching conditions between the interior and the exterior solutions are
obtained by the continuity of the metric and its derivatives along an
arbitrary hypersurface.
First of all we must establish the equality between the exterior and the interior metric at the surface of the star.
This can be done transforming the interior solution given in co-moving coordinates into the Schwarzschild frame.
The metric tensors in the two frames are related by the tensor transformation law:
\begin{equation}
\label{tensorlaw}
g'^{\mu \nu}=\frac{\partial x'^{\mu}}{\partial x^{\alpha}} \frac{\partial x'^{\nu}}{\partial x^{\beta}} \,g^{\alpha \beta}\,.
\end{equation}
The line element in the Schwarzschild frame is given by
\begin{equation}
ds^{2}=A^2\, c^{2} \,dT^{2} - B^2\, dR^{2} - 
R^{2} \,d \Omega^{2}\,,
\end{equation}
with $d\Omega=d \theta + {\rm sin}^{2} \theta\, d\phi$.
Using the Eq. (\ref{tensorlaw}) we obtain the non-trivial relations between the metric coefficients:
\begin{subequations}
\label{metriceqs}
\begin{eqnarray}
\frac{1}{A^2\,c^2}&=&T_{,t}^2\,\biggl(\frac{1}{a^2\,c^2}
\biggr)
-T_{,\mu}^2 \,\biggl(\frac{1}{b^2}\biggr)\,, \label{tensor1}\\ 
-\frac{1}{B^2}&=&R_{,t}^2\, \biggl(\frac{1}{a^2\,c^2}\biggr)-
R_{,\mu}^2 \,\biggl(\frac{1}{b^2}\biggr)\,, \label{tensor2}\\ 
0&=&R_{,t}\,T_{,t}\,\biggl(\frac{1}{a^2 c^2}\biggr)-
R_{,\mu}\,T_{,\mu}\,\biggl(\frac{1}{b^2}\biggr)\,. \label{tensor3}
\end{eqnarray}
\end{subequations}
 From Eq. (\ref{tensor2}) and using Eqs. (\ref{gammadef}) and (\ref{udefin}) we get:
\begin{equation}
\label{B1}
B^2=\bigl(\Gamma^2-u^2\bigr)^{-1}\,,
\end{equation}
or equivalently,
\begin{equation}
B=\biggl(1-\frac{2 m G}{R c^2} + \frac{G Q^2}{c^4 R^2}\biggr)^{-1/2}\,.
\end{equation}
From Eqs. (\ref{tensor1}) and (\ref{tensor3}):
\begin{equation}
\frac{1}{A^2}=\biggl(\frac{T_{,t}}{a}\biggr)^2 \biggl(\frac{\Gamma^2-u^2}{\Gamma^2}\biggr)\,.
\end{equation}
With a little amount of algebra we can transform this equation into:
$A^2 \,T_{,t}^2= B^2\, \Gamma^2\, a^2$, and using the Eq. (\ref{B1}):
\begin{equation}
\label{A1}
A^2 T_{,t}^2= B^2 a^2 \bigl(R_{,t}^2/a^2 c^2+B^{-2}\bigr)\,.
\end{equation}
This last equation can be obtained independently by comparing the induced metric, from interior and exterior solutions, 
on the hypersurface $\Sigma$: $t=t, \,\mu=\mu_s=constant,\,\theta=\theta,\phi=\phi$. The line element corresponding to the exterior solution is given by:
\begin{equation}
ds_+^2=A^2\, c^{2} \,dT^{2} - B^2\, dR^{2} - 
R_+^{2} \,d \Omega_+^{2}\,,
\end{equation} 
from now on a plus (minus) subscript or superscript denotes
exterior (interior) solutions.
The line element for the interior solution is (see Eq. \ref{comline}):
\begin{equation}
ds_-^2=a^2\, c^{2} \,dt^{2} - b^2\, d\mu^{2} - 
R_-^{2} \,d \Omega_-^{2}\,.
\end{equation} 
Due to  the spherical symmetry: $d \Omega_-=d \Omega_+$.
 The line element compatible with the induced  metric on the hypersurface, from the exterior solution  is\footnote{The  Poisson's book \cite{poisson} gives an excellent introduction to the theory of hypersurfaces in general relativity. The 
  matching conditions for the Oppenheimer-Snyder \cite{snyder} collapse are calculated there.}:
\begin{equation}
ds^2_{+|\Sigma}=(A^2\, c^{2} \,T^2_{,t} - B^2\, R^2_{,t})\, dt^{2} -R_+^{2} \,d \Omega_+^{2}\,,
\end{equation} 
and the metric induced on $\Sigma$ by the interior solution gives:
\begin{equation}
ds^2_{-|\Sigma}=a^2\, c^{2} \, dt^{2} -R_-^{2} \,d \Omega_-^{2}\,.
\end{equation} 
On $\Sigma$ we have $ds^2_{-|\Sigma}=ds^2_{+|\Sigma}$, so we obtain:
\begin{subequations}
\begin{eqnarray}
&&R_+=R_-\\
&&A^2\, c^{2} \,T^2_{,t} - B^2\, R^2_{,t}=a^2\, c^{2}\,. \label{A2}
\end{eqnarray} 
\end{subequations}
and we recovered the Eq. (\ref{A1}). So Eq. (\ref{A2}) (or equivalently Eq. \ref{A1}) are conditions for the continuity of the metric along $\Sigma$. From now on, we will use indistinctly $R=R^+=R^-$. 
The relation between the exterior and interior metric coefficients constitutes three equations with the four unknowns (see Eq. \ref{metriceqs}):
$R_{,t};\,R_{,\mu};\,T_{,t};\,T_{,\mu}$. Observe that $R_{,t}$ and $R_{,\mu}$ are obtained as  part of the interior solution.  In addition, the solution exterior to the sphere of matter must satisfy the field equations: $R_{\mu \nu}=0$, here $R_{\mu \nu}$ is the Ricci tensor. So we obtain (see for example \cite{weinberg} Pag. 180 and Pag. 337):
\begin{equation}
\label{A3}
A(R) \,c=\frac{1}{B(R)}\,.
\end{equation} 
The continuity of the metric is almost already established.
Only for completeness, and after a little algebra, we can find an equation for $T_{,\mu}$: $T_{,\mu}^2=u^2\, b^2\, B^4$ (although we will not use it).  

The equation of the hypersurface 
containing the surface of the star is 
\begin{equation}
\label{sigmas}
\Sigma_s: \,t=t,\,\mu=\mu_s,\,\theta=\theta,\,\phi=\phi\,,
\end{equation}
where $\mu_s$ is the Lagrangian mass at the surface of the star, so the metric coefficient $B$ is:
\begin{equation}
B=\biggl(1-\frac{2 M G}{R c^2} + \frac{G Q^2}{c^4 R^2}\biggr)^{-1/2}\,,
\end{equation}
with $M=m(t,\mu_s)$, $R=R(t,\mu_s)$, and $Q=Q(t,\mu_s)$.
The coefficient $A^2 c^2=g_{tt}$ is $A^2 c^2=1/B^2$, and we recover the Reissner-Nordstr\"om solution:
\begin{eqnarray}
ds^2&=&\biggl(1-\frac{2 M G}{R c^2} + \frac{G Q^2}{c^4 R^2}\biggr) dt^2
- \\
&&\biggl(1-\frac{2 M G}{R c^2} + \frac{G Q^2}{c^4 R^2}\biggr)^{-1} dR^2
 -R^2 d\theta^2 \nonumber \\ 
&& - R^2\,{\rm sin}^2 \theta\, d\phi^2\,.\nonumber
\end{eqnarray}
It is important to remark that we obtained the Reissner-Nordstr\"om solution by performing a tensor transformation of the interior solution. The exterior solution must satisfy the Einstein field equations, and so we used them to find the Eq. (\ref{A3}). Thus, the exterior and interior metric are consistently matched.

It remains to prove the continuity of the derivatives of the metric along $\Sigma_s$.  This is a more lengthy calculation and we will give only the the main algebraic steps and the final results.
The continuity of the derivatives of the metric is established from the continuity of the second fundamental form, or  extrinsic curvature tensor $K_{ab}$ of the hypersurface $\Sigma_s$. By definition (see \cite{poisson}, Pag. 59):
\begin{equation}
K_{ab}=n_{\alpha;\beta}\,e_a^{\alpha} e_b^{\beta}\,,
\end{equation}
here $n^{\alpha}$ is a unit vector normal to $\Sigma_s$; $e_a^{\alpha}=\partial x^{\alpha}/\partial y^{a}$, are basis vectors on $\Sigma_s$; the coordinates on $\Sigma_s$ are denoted with $y^a$, while $x^{\alpha}$ are the space-time coordinates. Equivalently :
$K_{ab}=\frac{1}{2}\bigl(\mathcal{L}_n g_{\alpha\beta}\bigr) e_a^{\alpha} e_b^{\beta}$, here $\mathcal{L}_n$ is a Lie derivative. 

We will choose $y^a=(t,\theta,\phi)$ as coordinates on $\Sigma_s$. The equation of the hypersurface approaching from the exterior is:
\begin{equation}
\Sigma_s^+ : T=T(t,\mu_s),\,R=R(t,\mu_s),\,\theta=\theta,\,\phi=\phi\,,
\end{equation}
The basis vectors on $\Sigma_s^+$ are:
\begin{subequations}
\begin{eqnarray}
&&e_{(t)}^{\alpha}=a^{-1}\,(T_{,t},R_{,t},0,0)\\
&&e_{(\theta)}^{\alpha}=(0,0,R^{-1},0)\\
&&e_{(\phi)}^{\alpha}=(0,0,0,R^{-1} {\rm sin}^{-1}\theta)\,,
\end{eqnarray}
\end{subequations}
in the coordinate basis $\mathcal{B}=\{\partial_T,\partial_R,\partial_\theta,\partial_\phi\}$,
and the normal co-vector is:
\begin{equation}
n_{\alpha}^+=a^{-1}\,(-R_{,t},T_{,t},0,0)\,.
\end{equation}
The equation for $\Sigma_s$ as seen from the interior is 
(see Eq. \ref{sigmas}):
\begin{equation}
\Sigma_s^- : t=t,\,\mu=\mu_s,\,\theta=\theta,\,\phi=\phi\,,
\end{equation}
the basis vectors on $\Sigma_s^-$ are:
\begin{subequations}
\begin{eqnarray}
&&e_{(t)}^{\alpha}=(a^{-1},0,0,0)\\
&&e_{(\theta)}^{\alpha}=(0,0,R^{-1},0)\\
&&e_{(\phi)}^{\alpha}=(0,0,0,R^{-1} {\rm sin}^{-1}\theta)\,,
\end{eqnarray}
\end{subequations}
in the coordinate basis $\mathcal{B'}=\{\partial_t,\partial_\mu,\partial_\theta,\partial_\phi\}$, and 
the normal co-vector is:
\begin{equation}
n_{\alpha}^-=(0,b,0,0)\,.
\end{equation}
If there are no surface distributions of energy-matter on $\Sigma_s$, it must be\footnote{See the Poisson's book \cite{poisson}, Pag. 69.}:
\begin{equation} 
\label{keq}
[K_{ab}] \equiv K^+_{ab}-K^-_{ab} = 0\,.
\end{equation}
 The 
angular components of this equation are 
$K^+_{\theta \theta}=K^-_{\theta \theta}$,
and $K^+_{\phi \phi}=K^-_{\phi \phi}$.
For $K^+_{\theta \theta}$ we have:
\begin{eqnarray}
\label{kp}
K^+_{\theta \theta}&=&\Gamma^R_{\theta \theta} \,n_{R}\,e_{(\theta)}^{\theta}\,e_{(\theta)}^{\theta}\\
&=&R^{-1}\,a^{-1}\,A\,c\,T_{,t}=R^{-1}\,\Gamma^+\,, \nonumber
\end{eqnarray}
here $\Gamma^\alpha_{\beta \gamma}$ is the connection. From now on, in the present section, we are not using the sum rule over repeated indexes. In the last two steps we used the Eqs. (\ref{A2}) and (\ref{A3}). 

For $K^-_{\theta \theta}$,
we find $K^-_{\theta \theta}=R_{,\mu} \,R^{-1}/b$, and using the Eqs. (\ref{gammadef})  and (\ref{gamma2}): $K^-_{\theta \theta}=R^{-1}\,\Gamma^-$, so Eq. (\ref{keq}) gives:
\begin{equation}
\Gamma^+(t,\mu_s)=\Gamma^-(t,\mu_s)\,,
\end{equation}
or equivalently:
\begin{eqnarray}
\label{gammaspm}
&&\biggl(1+\frac{u^2}{c^2}-\frac{2\,M\,G}{R\,c^2} + \frac{G\, Q^2}{c^4\, R^2}\biggr)^{-1/2}= \nonumber\\
&&\biggl(1+\frac{u^2}{c^2}-\frac{2\, m_s\, G}{R\, c^2} + \frac{G \,Q_s^2}{c^4\, R^2}\biggr)^{-1/2}\,.
\end{eqnarray}
Here $M=m(t,\mu_s+\epsilon)$ and $Q=Q(t,\mu_s+\epsilon)$, with $\epsilon$ a positive arbitrary  small number or zero.
The definition of $u$ assures its continuity across $\Sigma_s$, so $u^+=u^-$.
Moreover, we already 
found that $M$ and $Q$ are constants independent of $t$ and $\mu$,
so the Eq. (\ref{gammaspm}) implies that $\dot{m_s}=0$. Then,  using the Eq. (\ref{dmdt}) in co-movel coordinates ($\dot{Q}=0$ in both coordinate systems, see Eq. \ref{dqdt} and Eq. \ref{chargeconsintime}), we  obtain the boundary condition at the surface\footnote{This result is analogous to that obtained by D. L. Beckedorff, C. W. Misner and D. H. Sharp \cite{misner} for the non-charged case.}: 
$$P_s \equiv P(t,\mu_s)=0\,.$$
The equation  
$[K_{\phi \phi}]=0$ gives an analogous result.
  
The component $K^-_{tt}$ is:
\begin{eqnarray}
\label{kminus}
K^-_{tt}&=&-\Gamma_{tt}^\mu \,n_\mu \,e_{(t)}^t\, e_{(t)}^t=\,\nonumber \\
&&-\frac{1}{2}\,g^{\mu \mu} \frac{\partial{g_{tt}}}{\partial{\mu}}\,a^{-2}=
\frac{a_{\mu}}{a\,b}\,.
\end{eqnarray}

The external component is:
\begin{eqnarray}
K^+_{tt}&=&n_{t;t}\,e_{(t)}^t\,e_{(t)}^t
+n_{t;r}\,e_{(t)}^t\,e_{(t)}^r+ \nonumber\\
&& n_{r;t}\,e_{(t)}^r\,e_{(t)}^t
+n_{r;r}\,e_{(t)}^r\,e_{(t)}^r\,,
\end{eqnarray}
the sum rule is not used here. For clarity, we will define
$A^2\,c^2=B^{-2}=f$, so Eq. (\ref{A2}) becomes 
\begin{equation}
f\,\dot{T}^2-f^{-1}\,\dot{R}^2=a^2\,, 
\end{equation}
and an over-dot means $\partial_{t}$. Applying the over-dot operator to this equation we obtain an equation that will be used later:
\begin{equation}
\label{ddott}
\ddot{T}=\frac{2\,f^{-1}\,\dot{R}\,\ddot{R}-
2\,\dot{T}^2\,\dot{f}+2\,a\,\dot{a}+a^2\,f^{-1}\,\dot{f}}{2\,\dot{T}\,f}\,.
\end{equation}

So,
\begin{eqnarray}
K^+_{tt}&=&\{-\frac{\partial ( a^{-1}\,\dot{R} )}{\partial T}-\frac{1}{2}\, f \frac{\partial f}{\partial{R}}\,(a^{-1}\,\dot{T})\} \,a^{-2}\,\dot{T}^2+ \nonumber\\
&&\{-\frac{\partial ( a^{-1}\,\dot{R} )}{\partial R}+\frac{1}{2}\, f^{-1} \frac{\partial f}{\partial{R}}\,(a^{-1}\,\dot{R})\} \,a^{-2}\,\dot{T}\,\dot{R}\,+\nonumber\\
&&\{+\frac{\partial ( a^{-1}\,\dot{T} ) }{\partial T}+\frac{1}{2}\, f^{-1} \frac{\partial f}{\partial{R}}\,(a^{-1}\,\dot{R})\} \,a^{-2}\,\dot{T}\,\dot{R}\,+\nonumber\\
&&\{+\frac{\partial ( a^{-1}\,\dot{T} ) }{\partial R}- 
 \frac{1}{2}\, f \frac{\partial f^{-1}}{\partial{R}}\,(a^{-1}\,\dot{T})\} \,a^{-2}\,\dot{R}^2\,.\nonumber\\
\end{eqnarray}
Expanding this equation, and using the Eq. (\ref{ddott}) to simplify the result we obtain:
\begin{eqnarray}
K^+_{tt}=a^{-3} \,\biggl(\frac{2\,a\,\dot{a}\,\dot{R}^2-2\,a^2\,\dot{R}\,\ddot{R}-a^4\,\dot{f}}{2\,\dot{T}\,f\,\dot{R}}\biggr)\,,
\end{eqnarray}
further simplifications can be made using:
\begin{eqnarray}
\dot{f}&=&\frac{2\,m\,G}{R^2\,c^2}\,\dot{R}-\frac{2\,Q^2\,G}{R^3\,c^4}\,\dot{R}
-\frac{2\,\dot{m}\,G}{R\,c^2}\,,\\
\dot{u}&=&\frac{\ddot{R}}{a}-\frac{\dot{R}\,\dot{a}}{a^2}\,,\\
f\,\dot{T}&=&a\,(\dot{R}^2/a^2+f)^{1/2}=a\,\Gamma\,.
\end{eqnarray}
Thus we obtain:
\begin{eqnarray}
K^+_{tt}=\frac{1}{\Gamma}\,\biggr(-\frac{\dot{u}}{a\,c^2}-\frac{ m\,G}{R^2\,c^2}+\frac{G\,Q^2}{c^4\,R^3}
+\frac{G\,\dot{m}}{R\,\dot{R}\,c^2}\biggl)\,.
\end{eqnarray}
Comparing this equation with the Eq. (\ref{kminus})  we
obtain an expresion for $[K_{tt}]=0$:
\begin{equation}
\frac{1}{\Gamma}\,\biggr(-\frac{\dot{u}}{a}-\frac{ m\,G}{R^2}+\frac{G\,Q^2}{c^2\,R^3}
+\frac{G\,\dot{m}}{R\,\dot{R}}\biggl)=
\frac{a_{\mu}\,c^2}{a\,b}\,.
\end{equation}
Now rearranging and remembering that $b=1/4\,\pi\,\rho\,R^2$ and $G\,\dot{m}/\dot{R}=-4\,\pi\,G\,P\,R^2/c^2$,  using the Eq. (\ref{EM1}) to replace $a_{\mu}\,c^2/a$ we get:
\begin{eqnarray}
&&u_{t}=-a \biggr[4\,\pi\,R^2\,\frac{\Gamma}{w}\biggr(P_{,\mu}-\frac{Q\,Q_{\mu}}{4\,\pi\,R^4}\biggl)+\frac{G\,m}{R^2}+\nonumber\\
&&\frac{4\,\pi\,G}{c^2}\,P\,R-\frac{G\,Q^2}{c^2\,R^3}\biggl]\,.
\end{eqnarray}
Then, we see that the equation $[K_{tt}]=0$ is satisfied by virtue of the equation of motion  (see Eq. \ref{motion}), 
and thus
the exterior solution matches with the interior solution.  

\section{Equation of State\label{seceos}}
The equations obtained above must be supplemented
with an equation of state for the gas (EOS).
In particular we choose to represent the gas of neutrons as obeying the quantum statistic of a Fermion gas at zero temperature.
This let us use the Oppenheimer-Volkov numerical results
as a test-bed for the code. However, any other EOS can be 
equally implemented in the two codes presented in this paper. 

The energy per unit mass is given by \cite{shapiro}
\begin{equation}
\label{EOSe}
\delta\, c^2=\frac{m_n^4 c^5}{\pi^2 \hbar ^3}
\int_0^{x_F} \sqrt{x^2+1}\, x^2\,dx\,, 
\end{equation}
where $x=p/m_n c$ is the relativity parameter, and $p$ 
is the momentum of the particles.

The Fermi parameter is: $x_F=p_F/m_n c$, with $p_F=\frac{3 h^3}{8 \pi} n$, where $n$ is the number density of neutrons.
It can be written (we suppress the sub-index F from now on):
\begin{equation}
x=\biggl( \frac{\rho}{\rho_0}\biggr)^{1/3}\,,
\end{equation}
where 
\begin{equation}
\rho_0=\frac{m_n^4 c^3}{3 \pi^2 \hbar^3}=6.10656\times 10^{15}\,\,\,\, [{\rm g\,cm}^{-3}]\,.
\end{equation}
The Eq. (\ref{EOSe}) can be integrated to give:
\begin{eqnarray}
\label{EOSe2}
&&\delta\, c^2=\frac{m_n^4 c^5}{\hbar^3} 
\frac{1}{8 \pi^2} \biggl[ x \,\sqrt{1+x^2} \,\bigl(1+2 x^2\bigr)-\nonumber\\
&&\hspace{0.9cm}{\rm log} \bigl(x+\sqrt{1+x^2}\bigr) \biggr]\,, 
\end{eqnarray}
the quantity $\delta\, c^2$ is measured in [${\rm erg\, cm}^{-3}$], or equivalently in [${\rm dyn\, cm}^{-2}$].
This is an expression for the total energy of the gas, so it includes the rest mass 
energy density. Although this is obvious by definition (see Eq. \ref{EOSe}), it can be useful to check this expanding Eq. (\ref{EOSe2}) for $x\rightarrow0$ (or $x\rightarrow\infty$).
In this two cases
an split similar to Eq. (\ref{split}) is obtained. 

The pressure is given by \cite{shapiro}:
\begin{equation}
\label{press}
P=\frac{m_n^4 c^5}{3 \pi^2 \hbar^3} 
\int_0^{x_F} \frac{x^4}{\sqrt{x^2+1}}\, dx\,,
\end{equation}
and performing the integral:
\begin{eqnarray}
&&P=\frac{m_n^4 c^5}{\hbar^3} \frac{1}{8 \pi^2} 
\biggr[x\, \sqrt{1+x^2} \,\bigl(2 x^2/3-1\bigr) + \nonumber\\
&&\hspace{0.9cm}{\rm log} \bigl((x+\sqrt{1+x^2}\bigr) \biggr]\,, 
\end{eqnarray}
measured in [${\rm dyn\, cm}^{-2}$].

For charged matter, the equation of state must 
describe several species of particles, i.e.,
 neutrons, protons and electrons. However,
the amount of charge in all the models studied here is very small
to change the EOS. For  an extremal charged fluid ball there
are only one unpaired (not screened) proton over $10^{18}$  particles. In this case,
the change on the chemical potential of the neutrons 
is negligible and charge neutrality can be assumed from a microscopic
point of view \footnote{
Assuming charge neutrality in a gas of electrons, protons
and neutrons,
the chemical potential of the neutrons is
$$\mu_n=\mu_p+\mu_e.$$  Relaxing the condition of charge
neutrality, it is straightforward to show that for the case
of extremely charged matter, the chemical potential of the neutrons is
 $$\mu_n=\mu_p+\mu_e-10^{-18} \mu_e.$$ Thus, the EOS
is slightly modified, and charge neutrality can be assumed
from a microscopic point of view.
We will come back to this issue in a forthcoming paper.}.

The simulations were extended beyond the densities at which the EOS is valid. However, 
we want to take into account possible charge and pressure regeneration effects, that occurs at high densities.

\section{Numeric integration of the TOV
equations for charged stars.\label{sectov}}
The equations that must be
integrated to obtain the stars in hydrostatic equilibrium are conveniently written in dimensionless form.
From Eqs. (\ref{metrica0}) and (\ref{RE1}) it is possible to obtain an  equation for the metric component 
\mbox{$g_{\mu \mu}=b^2=e^{\lambda}$},
\begin{eqnarray}
\label{eqlambda}
\frac{d \lambda}{d R}= \frac{G \,Q^2}{c^4 R^3}\, e^{\lambda} 
+\frac{8 \pi G}{c^2} \,e^{\lambda}\, R\,\delta - \frac{e^{\lambda}-1}{R}\,.
\end{eqnarray}

In addition we use the chain rule on the TOV equation (Eq. \ref{TOVeq}):
\begin{eqnarray}
\label{TOVeq2}
&&\frac{d \rho}{dR}=-(\delta c^2 +P) \frac{\bigl(\frac{m G}{c^2}
+ \frac{4\pi G}{c^4}  P R^3-
\frac{G Q^2}{R c^4}\bigr)}
{R \bigl(R-\frac{2 m G}{c^2}+\frac{G Q^2}{R c^4}\bigr)} 
\frac{1}{(dP/d\rho)}+\nonumber\\
&&\hspace{0.9cm}\frac{Q}{4 \pi R^4} \frac{dQ}{dR} \frac{1}{(dP/d\rho)}\,.
\end{eqnarray}

The dimensionless equations (\ref{TOVeq2}) and (\ref{eqlambda})
 can be obtained with the replacements:
\begin{eqnarray*}
&&m=m_0 \,\bar{m}\,,\,\, P=P_0 \,\bar{P},\,\, Q=Q_0 \,\bar{Q}\, ,\mu=\mu_0 \,\bar{\mu}\\
&&R=R_0\,\bar{R}\,,\,\, \delta=\delta_0 \,\bar{\delta}\,,\,\,\rho_{ch}=\rho_{ch_0}\,
\bar{\rho_{ch}} \,.
\end{eqnarray*} 
Where the bar denotes dimensionless variables, and the subscript $``0"$ indicates the 
dimensional constants.

The full set of dimensionless equations to be integrated are:
\begin{subequations}
\begin{eqnarray}
\label{lambdaadim}
&&\frac{d {\lambda}}{d \bar{R}}= \frac{\bar{Q}^2}{ \bar{R}^3}\, e^{{\lambda}} 
+ e^{\lambda}\, \bar{R}\,\bar{\delta} - \frac{e^{\lambda}-1}{\bar{R}}\\
&&\frac{d\bar{Q}}{d \bar{R}}= \bar{\rho_{ch}}\,e^{\lambda/2}\,\bar{R}^2\\ 
&&\frac{d\bar{\mu}}{d \bar{R}}= \bar{\rho}\,e^{\lambda/2}\,\bar{R}^2\\ 
&&\label{TOVadim} \frac{d\bar{\rho}}{d\bar{R}}=-(\bar{\delta}+\bar{P}) \frac{\bigl(\bar{m}+  \bar{P} \bar{R}^3-
\frac{\bar{Q}^2}{\bar{R}}\bigr)}
{\bar{R} \bigl(\bar{R}-2 \bar{m} +\frac{\bar{Q}^2}{\bar{R}}\bigr)} 
\frac{1}{d\bar{P}/d\bar{\rho}}+\nonumber\\
&&\hspace{0.8cm}\frac{\bar{Q}}{\bar{R}^4} \frac{d\bar{Q}}{d\bar{R}} \frac{1}{d\bar{P}/d\bar{\rho}} \\
&&\label{massadim}
\frac{d \bar{m}}{d \bar{R}}=\bar{\delta}\, \bar{R}^2\, d \bar{R}
+\frac{\bar{Q}}{\bar{R}^2} \,\frac{d \bar{Q}}{d \bar{R}}\,.
\end{eqnarray}
\end{subequations}

The dimensional constants are given by the set of
equations:
\begin{subequations}
\begin{eqnarray}
&&M_0= 4 \,\pi\, \rho_0 \,R_0^3=
\frac{c^3}{\sqrt{4 \,\pi \,\rho_0 \,G^3}}\\
&&\mu_0=M_0 \\
&&Q_0=\sqrt{G}\,M_0\\
&&R_0= \frac{M_0 \,G}{c^2}=\frac{c}{\sqrt{4\, \pi\, \rho_0\, G}} \\
&&P_0=\rho_0 \,c^2\\
&&\rho_{ch_0}=\rho_0\\
&&\delta_0=\rho_0\,.
\end{eqnarray}
\end{subequations}

It is natural to choose:
\begin{equation} 
\rho_0=\frac{m_n^4\,c^3}{\hbar^3}=1.80808\times 10^{17}\,\,\,{\rm g \,cm}^{-3}\,,
\end{equation}
since $\rho_0\,c^2$ is the factor in the expressions for
the pressure (Eq. \ref{press}) and energy (Eq. \ref{EOSe}).
Hence:
\begin{subequations}
\begin{eqnarray}
&&M_0=1.03706\times10^{33}\,\,\,\,{\rm g}\\
&&R_0=7.69944\times10^4\,\,\,\,{\rm cm}\\
&&\rho_{ch_0}=1.80808\times 10^{17}\,\,\,{\rm g \,cm}^{-3}\\
&&\rho_0\,c^2=1.62502\times10^{38}\,\,\,\,{\rm dyn\, cm}^{-2}\\
&&Q_0=2.67888\times10^{29}\,\,\,\,{\rm StatCoulombs}\,.
\end{eqnarray}
\end{subequations}

Moreover the set of equations (\ref{lambdaadim})-(\ref{massadim}) are invariant under the
transformations: $R_0 \rightarrow R_0\,\alpha$,
$M_0 \rightarrow M_0\,\alpha$, and 
$\rho_0 \rightarrow \rho_0/\alpha^2$, etc (the other constants
can be obtained from these), with $\alpha$
an arbitrary number different from zero. So, 
it is possible to choose any other set of constants
consistent with these transformations. In particular,
the dimensional constants obtained by Oppenheimer and Volkov 
\cite{oppen}, are recovered by setting: $\alpha=\sqrt{32 \pi^2}$. In this
case: $R_0=1.36831 \times 10^6$ cm; $M_0=1.84302\times10^{34}$ g, etc, 
equivalent to the constants obtained in that paper \cite{oppen}.

In the Eq. (\ref{TOVadim}) the term $d\bar{P}/d\bar{R}$,
is obtained deriving the Eq. (\ref{press}):
\begin{equation}
\frac{d \bar{P}}{d \bar{R}}=\frac{x^2}{3 \pi^2 \sqrt{1+x^2}}\,.
\end{equation}

It is  used a $4^{th}$ order Runge-Kutta scheme
to integrate simultaneously the set of equations (\ref{lambdaadim})-(\ref{massadim}).

The charge distribution is chosen
proportional to the rest mass distribution: 
$\rho_{ch}=\alpha(\mu,t)\, \rho$. For simplicity,
in this paper we will concentrate on 
the case $\alpha=constant$.

The code implemented to integrate the equations above
is called HE05v1.
The HE05v1 is used to build neutron star models in hydrostatic equilibrium.
 This models constitutes  an  initial data set
for another code presented below. 

\section{Numerical integration of the Eintein-maxwell equations\label{sechydro}}
In this section we collect the equations that must be integrated numerically in order to simulate the temporal evolution of the 
stellar models constructed with the HE05v1 code. 
The set of equations is:
\begin{subequations}
\label{allmotion}
\begin{eqnarray}
\noindent
&&du=-a \biggl[4 \,\pi \,R^2\, \frac{\Gamma}{w} \biggl(P_{,\mu} - \frac{Q\,Q_{,\mu}}{4\, \pi \,R^4}\biggr)
+\frac{G\,m}{R^2}+ \label{ME1} \nonumber \label{dynam}\\
&&\hspace{0.8cm}\frac{4\,\pi\, G}{c^2}\,P\, R-\frac{G\, Q^2}{c^2\, R^3} \biggr]\,dt \hspace{1.56cm}[\ref{motion}]\\
&&w=1+\frac{\epsilon}{c^2}+\frac{P}{\rho\, c^2} \hspace{2.96cm}[\ref{enthalpy}] \\
&&\Gamma^2=1+\frac{u^2}{c^2}-\frac{2 m G}{R c^2} + \frac{G Q^2}{c^4 R^2} \hspace{1.335cm}[\ref{GammaHE}] \\
&&d\mu=4 \,\pi\, \rho\, R^2\, dR/\Gamma\hspace{2.8cm}[\ref{gammadef}] \\
&&{a w}={a_0 w_0}\,{\rm exp}\,\biggl[\,\int_0^\mu\,\biggl( d\epsilon + P \,d \biggl(\frac{1}{\rho} \biggr)+ \nonumber\\
&& \hspace{0.8cm} \frac{Q \,d Q}{4\, \pi\, R^4\, \rho} 
\biggr) /w\,c^2\biggr] \hspace{2.45cm}[\ref{gtt2}]\\
&&dm=4\, \pi\,  \rho\, \biggl(1+\frac{\epsilon}{c^2} \biggr) \,R^2\, R_{,\mu}\, d\mu + \nonumber \label{hamilton}\\ 
&&\hspace{0.8cm} \frac{1}{c^2} \frac{Q Q_{,\mu}}{R} d\mu  \hspace{3.25cm}[\ref{totalmass}]\\ 
&&{\rho\,R^2}={\rho_0\, R_0^2}\,\,{\rm exp}\,
\biggr[-\int_0^t a \frac{u_{,\mu}}{R_{,\mu}} \,dt\biggl] \label{momentum}
\hspace{0.57cm}[\ref{massconservation}]\\
&&d\epsilon=-P\, d\biggl(\frac{1}{\rho}\biggr) \hspace{3.39cm}[\ref{1stlaw}]\\
&&dQ=4\, \pi\, \rho_{ch} \,R^2\, dR/\Gamma \label{EGamma}\\ 
&&d{\mu}=4 \,\pi\, \rho\,{R}^2\,dR/\Gamma\,.\hspace{2.7cm}[\ref{gammadef}] \label{Emu}
\end{eqnarray}
\end{subequations}
Where we mean $a_0\,w_0=(a\,w)(t,0)$ and $P_0\,R^2_0=(P\,R^2)(0,\mu)$. The references to other equations are enclosed in square brackets.
Comparing the Eqs. (\ref{allmotion})  with the May and White equations \cite{may}, 
it can be seen that they share a very similar structure when $Q=0,\,\,\,\forall\,(\mu,t)$.
We have chosen a charge distribution proportional to the rest mass distribution so we obtain the Eq. (\ref{EGamma}).
The Eq. (\ref{dynam}) is the only equation that have a second time derivative and constitutes the dynamical part of
the Einstein-Maxwell equations. The Eqs. (\ref{hamilton}) and (\ref{momentum}) represent the Hamiltonian and momentum constraint equations, respectively \footnote{See Ref. \cite{weinberg}, Pag. 164.}. The numerical code to integrate the equations above are called \mbox{Collapse05v3}.

The initial conditions at $t=t_0$ are obtained with the HE05v1 for each stellar model,
and consist on: the initial mass density distribution $\rho(\mu,t_0)$; the initial electric charge density distribution $\rho_{ch}(\mu,t_0)$; the
 kinetic energy $\epsilon(\mu,t_0)$; the cell spacing $dR(\mu,t_0)$ and the surface radius $R(\mu_{s},t_0)$. Here, $\mu_{s}$ is the mass coordinate of the surface.
Then, the output of the HE05v1 is used as an initial Cauchy
data set in the space-time hypersurface $t=t_0$, $\mu=\mu$, $\theta=\theta$, $\phi=\phi$, for the \mbox{Collapse05v3}.

The boundary conditions are:
\begin{subequations}
\begin{eqnarray}
&&P=0, \,\,\,{\rm at}\,\,\mu=\mu_{s},\,\,\,\,\forall \,t \label{BCp}\\ 
&&a=1,\,\,\,\,{\rm at}\,\,\mu=\mu_{s},\,\,\,\,\forall \,t \label{BCa}\\ 
&&u=0, \,\,\,\,{\rm at}\,\,t=t_0,\,\,\, \forall \,\mu \\ 
&&R=0,\,\,\,{\rm at}\,\,\,\mu=0, \,\,\,\forall \,t\,. \label{BCr}
\end{eqnarray}
\end{subequations}
The condition given in Eq. (\ref{BCa}), makes the time coordinate synchronized with an observer co-moving with the surface of the star. The conditions expressed in the Eqs. (\ref{BCp})-(\ref{BCr}), also gives:
\begin{eqnarray*}
&&\Gamma=1,\,Q=0,\,m=0,\,\,\,{\rm at}\,\,\mu=0,\,\,\,\,\forall \,t \\ 
\end{eqnarray*}

The finite difference algorithm was implemented using
 a method similar to that developed by May and White \cite{may}.
The Eqs. (\ref{ME1})-(\ref{Emu}) were integrated with a leap-frog finite 
difference method plus a predictor-corrector step. 
We call this numerical code as \mbox{Collapse05v3}. The integration
is iterated according to a desired error control. 
The method is second order accurate in space and 
time \footnote{See the article of Bernstein et al. in  Ref. \cite{evans}, Pag 57.}.
Each experiment was repeated with different number of particles and with different values of the numerical 
viscosity parameter, in order to check the convergence of the results.
In general, good results are obtained using $\sim 200$ points along the coordinate $\mu$. 
The compatibility between the two codes is an
indirect check of the convergence properties of the \mbox{Collapse05v3}, since the HE05v1 is $4th$ order accurate. For example, at $t=t_0$ there  exists
 differences in the integrated mass between the two codes of the order of $\sim 0.01\,M_{\odot}$, using $200$ points in the \mbox{Collapse05v3}. This difference can be made $< 0.001\,M_{\odot}$ by taking, instead, $800$ points, and so on.  
For the integration in the time dimension is needed to take initially small time-steps $dt=10^{-5}$ s, but the code is capable of  self-adjust
the time-step so as to take small changes in the physical variables. Therefore, with this algorithm it is indifferent to take an initial time-step of $10^{-3}$ s or $10^{-5}$ s, the code always self-adjust the time-step and prove to converge always to the same solution.
The shocks are treated by adding an artificial viscosity, which is
non-zero only on discontinuities. 
Its effects are to smear out
the discontinuities over several cells, and  to reduce the post-shock oscillations and the numerical errors related to the leap-frog method. We experienced with distinct functional forms for the viscosity term, probing the method of May and White \cite{may} to be the best in the present algorithm.   However, there are not strong shocks (like in Ref. \cite{ghezzi2}) in the set of simulations studied in this paper and with the particular initial conditions chosen.

In the \mbox{Collapse05v3} the Eqs. (\ref{EGamma}) and (\ref{Emu})
are not integrated in time, and the rest mass and charge are constrained to be constant in the layers of matter (as it must be). We check that this kind of algorithm reduce the numerical errors.  

As was showed in the Sec. (\ref{relatequ})
the total mass-energy at the surface of the star is also a constant $m(t,\mu_{sup})=M=constant$. The numerical errors have an impact on the constancy of the total mass-energy $M$, and this is the physical quantity with which is  checked the accuracy of the simulation.
In all the runs  $M$ is very well conserved, 
with the exception at the time when a trapped surface forms. In that case, very large gradients forms and serious numerical instabilities appear. Although,  it is possible to manage the code to keep the error in mass-energy conservation below $\sim 5 \,\%$ at the time the apparent horizon forms. 
Of course, once the trapped surface forms nothing can change the fate of the matter inside the apparent horizon (see the Sec. \ref{subsubsec:AH}).
The simulation must be discontinued soon after the matter
cross the gravitational radius because the time-step attains
very small values. 
In general, when $dt$ is 
roughly $dt<10^{-10}$ s, it
is needed a very large number of time-steps to obtain further progress and this consume too much computer time. 
The reason is that the large gradients
produce huge changes in the physical variables and a very small
time-step is required to keep the errors at a low level. 
Thus, it is not possible to follow the dynamical evolution  until a stationary regime is reached (or the formation of an event horizon, if ever possible). When the outer
trapped surface forms, some layers of matter external to it can be collapsing or expanding and is not possible in general to say which will be their fate. However, the matter that falls in the trapped surface can not escape from it and this is enough to say that the star (or some part of it)  collapsed.

There are two different sources of perturbation introduced in  the models:
the first source of perturbation is of numerical origin and cames from the mapping of the equilibrium models in the evolution
code. This mapping is not exact, and there is a difference in the integrated rest mass and total mass between the two codes that is taken
  $\le 0.01\,M_{\odot}$. This difference can be made as small as wanted
by taking a higher number of points to perform the integration with the \mbox{Collapse05v3}; the second source of perturbation, is introduced by multiplying the kinetic energy
by a small constant grater than one, i.e., $\epsilon \rightarrow \alpha\, \epsilon$. In the present simulations we choose $\alpha$ to take the values $\alpha \sim 1.01-1.04$.

The structure of the space-time of a collapsed charged star is very complex, with the possible existence of time-like singularities and tunnels connecting several disjoint asymptotically flat space-time regions. This tunnels and the true singularities, pertains to the analytically extended portions of the space-time, and the present code is not prepared to reach that regions of the total manifold. 
In fact, with the present code is not possible to reach the Cauchy horizon where important physical effects must occur 
\cite{poisson}, \cite{piran}. 
However, from an astrophysical point of view
is interesting  to know what happens outside the apparent horizon, where the astronomers live.

We will not describe the numerical methods in more detail
because is not the objective of this paper. 
Moreover, the numerical methods are very well known 
and better explained in textbooks and papers (see for example Ref. \cite{evans}).
  The \mbox{Collapse05v3} is an extended version of a preliminary code developed by Ghezzi (2003), earlier  test-beds and results obtained with this code were published in collaboration 
\cite{ghezzi}, \cite{ghezzi2}.

\begin{table}
\caption{\label{tab:table1}Mass, radius, and central density
for neutron star models with zero charge.}

\begin{ruledtabular}
\begin{tabular}{rrcccccc}
Model& Radius\footnotemark[1]  & Mass\footnotemark[2] & Density\footnotemark[3]  \\
&[km]&$M_{\odot}$ & [${\rm g\,cm}^{-3}$]  \\ 
\hline
  1 &    46.735 &      0.035 &  $ 1.000\times\,10^{12} $ \\
  2 &    43.117 &      0.044 &  $ 1.589\times\,10^{12} $ \\
  3 &    39.579 &      0.055 &  $ 2.525\times\,10^{12} $ \\
  4 &    37.297 &      0.069 &  $ 4.013\times\,10^{12} $ \\
  5 &    33.892 &      0.087 &  $ 6.377\times\,10^{12} $ \\
  6 &    31.526 &      0.109 &  $ 1.013\times\,10^{13} $ \\
  7 &    29.143 &      0.135 &  $ 1.610\times\,10^{13} $ \\
  8 &    26.796 &      0.168 &  $ 2.559\times\,10^{13} $ \\
  9 &    24.525 &      0.208 &  $ 4.067\times\,10^{13} $ \\
 10 &    22.801 &      0.255 &  $ 6.463\times\,10^{13} $ \\
 11 &    20.686 &      0.309 &  $ 1.027\times\,10^{14} $ \\
 12 &    19.037 &      0.372 &  $ 1.632\times\,10^{14} $ \\
 13 &    17.156 &      0.439 &  $ 2.593\times\,10^{14} $ \\
 14 &    15.664 &      0.510 &  $ 4.121\times\,10^{14} $ \\
 15 &    14.042 &      0.579 &  $ 6.549\times\,10^{14} $ \\
 16 &    12.739 &      0.640 &  $ 1.041\times\,10^{15} $ \\
 17 &    11.220 &      0.686 &  $ 1.654\times\,10^{15} $ \\
 18 &    10.004 &      0.712 &  $ 2.628\times\,10^{15} $ \\
 $19\footnotemark[4]\hspace{-0.145cm}$ & 8.795 &  0.715 &  $4.177\times\,10^{15}$ \\
 20 &     7.823 &      0.696 &  $ 6.637\times\,10^{15} $ \\
 21 &     6.867 &      0.658 &  $ 1.055\times\,10^{16} $ \\
 22 &     6.164 &      0.608 &  $ 1.676\times\,10^{16} $ \\
 23 &     5.582 &      0.552 &  $ 2.663\times\,10^{16} $ \\
 24 &     5.195 &      0.496 &  $ 4.232\times\,10^{16} $ \\
 25 &     4.980 &      0.446 &  $ 6.726\times\,10^{16} $ \\
 26 &     4.984 &      0.406 &  $ 1.069\times\,10^{17} $ \\
 27 &     5.209 &      0.380 &  $ 1.698\times\,10^{17} $ \\
 28 &     5.627 &      0.371 &  $ 2.699\times\,10^{17} $ \\
 29 &     6.128 &      0.379 &  $ 4.289\times\,10^{17} $ \\
 30 &     6.493 &      0.398 &  $ 6.816\times\,10^{17} $ \\
 31 &     6.627 &      0.420 &  $ 1.083\times\,10^{18} $ \\
 32 &     6.605 &      0.436 &  $ 1.721\times\,10^{18} $ \\
 33 &     6.480 &      0.444 &  $ 2.735\times\,10^{18} $ \\
 $34\footnotemark[5]\hspace{-0.135cm}$ &     6.343 &      0.445 &  $4.347\times\,10^{18}$ \\
 35 &     6.225 &      0.441 &  $ 6.907\times\,10^{18} $ \\
 36 &     6.148 &      0.436 &  $ 1.098\times\,10^{19} $ \\
 37 &     6.118 &      0.431 &  $ 1.744\times\,10^{19} $ \\
 38 &     6.124 &      0.427 &  $ 2.772\times\,10^{19} $ \\
 39 &     6.151 &      0.425 &  $ 4.405\times\,10^{19} $ \\
 40 &     6.183 &      0.424 &  $ 7.000\times\,10^{19} $ \\
\end{tabular}
\end{ruledtabular}
\footnotetext[1]{Radius at the surface of the star measured in [km].}
\footnotetext[2]{Total mass-energy (Eq. \ref{totalmass})
 for each stellar model given 
in solar masses.}
\footnotetext[3]{Central density of each model measured
in [${\rm g\,cm}^{-3}$].}
\footnotetext[4]{This is the maximum neutron stars mass, 
calculated with mass steps of $0.0015\,M_{\odot}$.
It is also the first maximum indicated in the Fig. (\ref{fig:massadens}b).}
\footnotetext[5] {This is the second maximum indicated in the Fig. (\ref{fig:massadens}b)}

\end{table}

\begin{table}
\caption{\label{tab:table2}Mass, radius, and central density
for neutron star models with $Q/\sqrt{G}\mu=0.8$.}

\begin{ruledtabular}
\begin{tabular}{rrcccccc}
Model& Radius\footnotemark[1]  & Mass\footnotemark[2] & Density\footnotemark[3]  \\
&[km]&$M_{\odot}$ & [${\rm g\,cm}^{-3}$]  \\ 
\hline
  1 &              78.965 &               0.165 & $1.000\times\,10^{12}$  \\
  2 &              72.271 &               0.207 & $1.589\times\,10^{12}$  \\
  3 &              67.192 &               0.259 & $2.525\times\,10^{12}$  \\
  4 &              62.086 &               0.323 & $4.013\times\,10^{12}$  \\
  5 &              57.066 &               0.402 & $6.377\times\,10^{12}$  \\
  6 &              52.212 &               0.497 & $1.013\times\,10^{13}$  \\
  7 &              48.287 &               0.610 & $1.610\times\,10^{13}$  \\
  8 &              43.810 &               0.743 & $2.559\times\,10^{13}$  \\
  9 &              40.145 &               0.895 & $4.067\times\,10^{13}$  \\
 10 &              36.633 &               1.062 & $6.463\times\,10^{13}$  \\
 11 &              32.922 &               1.239 & $1.027\times\,10^{14}$  \\
 12 &              29.522 &               1.414 & $1.632\times\,10^{14}$  \\
 13 &              26.422 &               1.573 & $2.593\times\,10^{14}$  \\
 14 &              23.604 &               1.700 & $4.121\times\,10^{14}$  \\
 15 &              20.640 &               1.780 & $6.549\times\,10^{14}$  \\
 $16\footnotemark[4]\hspace{-0.145cm}$ &  18.217 &               1.803 & $1.041\times\,10^{15}$  \\
 17 &              15.914 &               1.767 & $1.654\times\,10^{15}$  \\
 18 &              13.896 &               1.679 & $2.628\times\,10^{15}$  \\
 19 &              12.131 &               1.552 & $4.177\times\,10^{15}$  \\
 20 &              10.681 &               1.402 & $6.637\times\,10^{15}$  \\
 21 &               $9.643$ &               1.245 & $1.055\times\,10^{16}$  \\
 22 &               $8.893$ &               1.097 & $1.676\times\,10^{16}$  \\
 23 &               $8.460$ &               0.966 & $2.663\times\,10^{16}$  \\
 24 &               $8.380$ &               0.862 & $4.232\times\,10^{16}$  \\
 25 &               $8.810$ &               0.793 & $6.726\times\,10^{16}$  \\
 26 &               $9.663$ &               0.766 & $1.069\times\,10^{17}$  \\
 27 &              10.802 &               0.785 & $1.698\times\,10^{17}$  \\
 28 &              11.695 &               0.841 & $2.699\times\,10^{17}$  \\
 29 &              12.088 &               0.904 & $4.289\times\,10^{17}$  \\
 30 &              12.047 &               0.951 & $6.816\times\,10^{17}$  \\
 31 &              11.753 &               0.973 & $1.083\times\,10^{18}$  \\
 $32\footnotemark[5]\hspace{-0.135cm}$ &              11.431 &               0.975 & $1.721\times\,10^{18}$  \\
 33 &              11.166 &               0.964 & $2.735\times\,10^{18}$  \\
 34 &              10.984 &               0.948 & $4.347\times\,10^{18}$  \\
 35 &              10.907 &               0.933 & $6.907\times\,10^{18}$  \\
 36 &              10.910 &               0.922 & $1.098\times\,10^{19}$  \\
 37 &              10.972 &               0.916 & $1.744\times\,10^{19}$  \\
 38 &              11.047 &               0.916 & $2.772\times\,10^{19}$  \\
 39 &              11.119 &               0.918 & $4.405\times\,10^{19}$  \\
 40 &              11.171 &               0.922 & $7.000\times\,10^{19}$  \\
\end{tabular}
\end{ruledtabular}
\footnotetext[1]{Radius at the surface of the star measured in [km].}
\footnotetext[2]{Total mass-energy (Eq. \ref{totalmass})
 for each stellar model given 
in solar masses.}
\footnotetext[3]{Central density of each model measured
in [${\rm g\,cm}^{-3}$].}
\footnotetext[4]{This model corresponds to the first mass maximum.}
\footnotetext[5]{This model corresponds to the second mass maximum.}

\end{table}

\begin{table}
\caption{\label{tab:table3}Mass, radius, and central density
for neutron star models with $Q/\sqrt{G}\mu=0.97$.}

\begin{ruledtabular}
\begin{tabular}{rrcccccc}
Model& Radius\footnotemark[1]  & Mass\footnotemark[2] & Density\footnotemark[3]  \\
&[km]&$M_{\odot}$ & [${\rm g\,cm}^{-3}$]  \\ 
\hline
  1 &             204.307 &             2.964 &  $ 1.000\times\,10^{12} $\\
  2 &             187.353 &             3.606 &  $ 1.589\times\,10^{12}  $ \\
  3 &             171.069 &             4.338 &  $ 2.525\times\,10^{12}  $ \\
  4 &             154.482 &             5.144 &  $ 4.013\times\,10^{12}  $ \\
  5 &             140.106 &             5.991 &  $ 6.377\times\,10^{12}  $ \\
  6 &             125.026 &             6.830 &  $ 1.013\times\,10^{13}  $ \\
  7 &             111.394 &             7.594 &  $ 1.610\times\,10^{13}  $ \\
  8 &              98.495 &             8.208 &  $ 2.559\times\,10^{13}  $ \\
  9 &              85.966 &             8.603 &  $ 4.067\times\,10^{13}  $ \\
 $10\footnotemark[4]\hspace{-0.145cm}$ &              75.006 &             8.734 &  $ 6.463\times\,10^{13}  $ \\
 11 &              64.658 &             8.590 &  $ 1.027\times\,10^{14}  $ \\
 12 &              55.735 &             8.194 &  $ 1.632\times\,10^{14}  $ \\
 13 &              47.761 &             7.599 &  $ 2.593\times\,10^{14}  $ \\
 14 &              40.928 &             6.874 &  $ 4.121\times\,10^{14}  $ \\
 15 &              35.073 &             6.085 &  $ 6.549\times\,10^{14}  $ \\
 16 &              30.232 &             5.293 &  $ 1.041\times\,10^{15}  $ \\
 17 &              26.512 &             4.541 &  $ 1.654\times\,10^{15}  $ \\
 18 &              23.498 &             3.861 &  $ 2.628\times\,10^{15}  $ \\
 19 &              21.359 &             3.272 &  $ 4.177\times\,10^{15}  $ \\
 20 &              20.114 &             2.782 &  $ 6.637\times\,10^{15}  $ \\
 21 &              19.849 &             2.399 &  $ 1.055\times\,10^{16}  $ \\
 22 &              20.858 &             2.135 &  $ 1.676\times\,10^{16}  $ \\
 23 &              23.870 &             2.018 &  $ 2.663\times\,10^{16}  $ \\
 24 &              29.138 &             2.102 &  $ 4.232\times\,10^{16}  $ \\
 25 &              35.184 &             2.420 &  $ 6.726\times\,10^{16}  $ \\
 26 &              38.527 &             2.835 &  $ 1.069\times\,10^{17}  $ \\
 27 &              38.479 &             3.128 &  $ 1.698\times\,10^{17}  $ \\
 $28\footnotemark[5]\hspace{-0.135cm}$ &              36.825 &             3.236 &  $ 2.699\times\,10^{17}  $ \\
 29 &              34.905 &             3.210 &  $ 4.289\times\,10^{17}  $ \\
 30 &              33.329 &             3.120 &  $ 6.816\times\,10^{17}  $ \\
 31 &              32.310 &             3.014 &  $ 1.083\times\,10^{18}  $ \\
 32 &              31.826 &             2.923 &  $ 1.721\times\,10^{18}  $ \\
 33 &              31.805 &             2.861 &  $ 2.735\times\,10^{18}  $ \\
 34 &              32.105 &             2.833 &  $ 4.347\times\,10^{18}  $ \\
 35 &              32.551 &             2.833 &  $ 6.907\times\,10^{18}  $ \\
 36 &              32.978 &             2.853 &  $ 1.098\times\,10^{19}  $ \\
 37 &              33.280 &             2.880 &  $ 1.744\times\,10^{19}  $ \\
 38 &              33.423 &             2.904 &  $ 2.772\times\,10^{19}  $ \\
 39 &              33.423 &             2.921 &  $ 4.405\times\,10^{19}  $ \\
 40 &              33.342 &             2.928 &  $ 7.000\times\,10^{19} $ \\
\end{tabular}
\end{ruledtabular}

\footnotetext[1]{Radius at the surface of the star measured in [km].}
\footnotetext[2]{Total mass-energy (Eq. \ref{totalmass})
 for each stellar model given 
in solar masses.}
\footnotetext[3]{Central density of each models measured
in [${\rm g\,cm}^{-3}$].}
\footnotetext[4]{This model corresponds to the first mass maximum.}
\footnotetext[5]{This model corresponds to the second mass maximum.}

\end{table}

\section{Results\label{secresults}}

\subsection{Charged neutron stars in hydrostatic equilibrium}
\begin{figure*}[h]
\begin{center}
$\begin{array}{c@{\hspace{1in}}c}
\multicolumn{1}{l}{\mbox{\bf (a)}} &
\multicolumn{1}{l}{\mbox{\bf (b)}} \\ [-0.53cm]
\epsfig{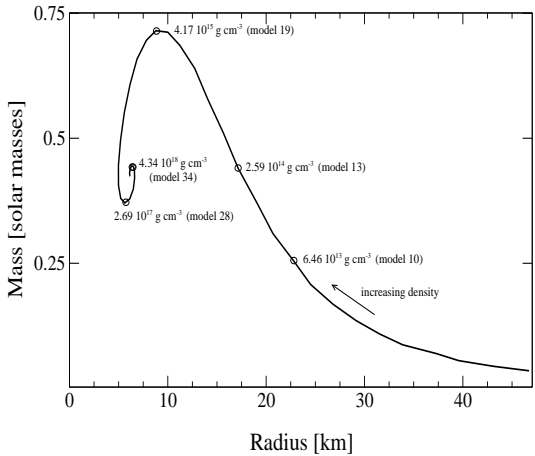} &
\epsfig{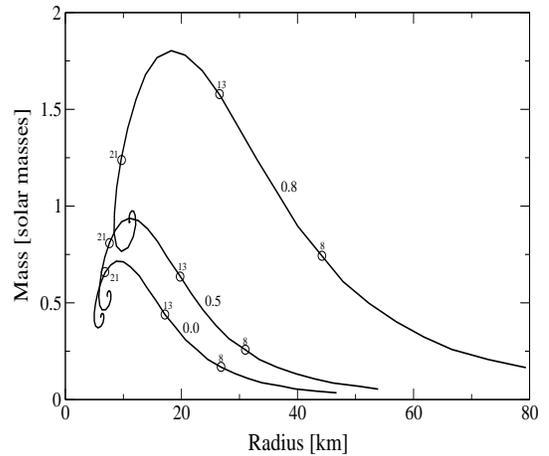} 
\\ [0.4cm]
\mbox{\bf (a)} & \mbox{\bf (b)}
\end{array}$
\end{center}
\caption{Mass-radius relation for neutron stars with zero charge (Fig. a), and mass-radius relation for neutron stars with different amounts of total
electric charge. Each curve is labeled by the charge to mass ratio of the stars: $Q/\sqrt{G} \mu=0,\,\,0.5,\,\,0.8$. (Fig. b).}
\label{fig:massarad}
\end{figure*}

The Fig. (\ref{fig:massarad}a) shows the mass-radius relation
for neutron stars with zero charge. 
In this curve different points  correspond to stars with different central densities and total number
 of nucleons. 
The results 
are in  agreement with the results of
Oppenheimer and Volkov \cite{oppen}.  

The Fig. (\ref{fig:massarad}b) shows the mass-radius relations for
models with different values of the charge to mass ratio
$Q/\sqrt{G} \mu$. 

In the Figs. (\ref{fig:massarad}) the central density 
of the models
increase from right to left, and following counterclockwise
sense inside the spiral.

\begin{figure*}[h]
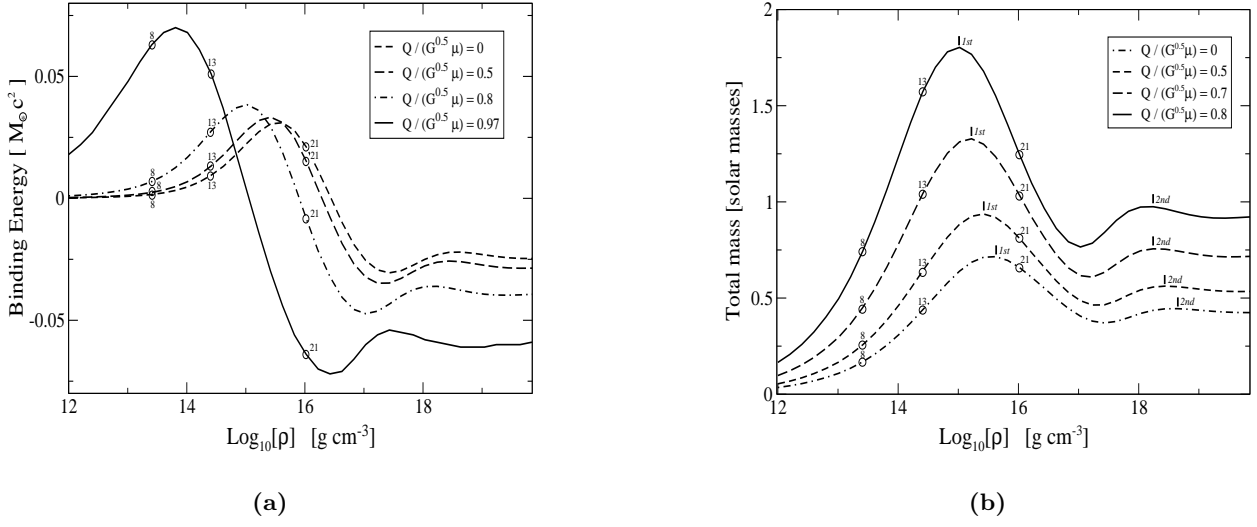

\begin{center}
$\begin{array}{c@{\hspace{1in}}c}
\multicolumn{1}{l}{\mbox{\bf (a)}} &
\multicolumn{1}{l}{\mbox{\bf (b)}} \\ [-0.53cm]
\epsfig{width=7cm,height=6cm, file=PRDIIbindingdens.eps} &
\epsfig{width=7cm,height=6cm, file=PRDIImassadens3.eps}
\\ [0.4cm]
\mbox{\bf (a)} & \mbox{\bf (b)}
\end{array}$
\end{center}
\caption{Total binding energy  in units of $M_{\odot} c^2$, for models with different values
of the charge to mass ratio $Q/\sqrt{G} \mu=0,\,\,0.5,\,\,0.8,\,\,0.97$ (Fig. a), and mass {\it versus} central density for models with different charge to mass ratio  $Q/\sqrt{G} \mu=0,\,\,0.5,\,\,0.7,\,\,0.8$ (Fig. b).}
\label{fig:massadens}
\hspace{3.6cm}
\end{figure*}

The factor
$Q/\sqrt{G} \mu$ is a constant in each  curve of the Fig. (\ref{fig:massarad}b), but $Q/\sqrt{G} M$ varies along   them. 
This is because the binding energy $(\mu-M) c^2$
is not constant along a curve with constant $Q/\sqrt{G} \mu$. In other words, the binding energy
varies with  the central density of the model. This can be seen in the  Fig. (\ref{fig:massadens}a), which in addition shows that the binding energy vary as a function of the charge, as well. The total binding energy of the stars increase with its charge  (see Sub-sec. \ref{subsec:bindenerg}). For example, the maximum mass model with $Q/\sqrt{G}\mu=0.97$, has a binding energy larger than the maximum mass model with $Q/\sqrt{G}\mu=0.8$. Moreover, in each curve,
the maximum is attained at lower densities for higher total charge (see Fig. \ref{fig:massadens}a). 

In the Fig. (\ref{fig:massadens}b) can be seen that for  higher values of the total charge (higher $Q/\sqrt{G} \mu$) the maximum mass and the radius of the models became larger.

The numbered circles in the Figs. (\ref{fig:massarad}) and (\ref{fig:massadens}), indicates
some stars that have the same central density. This models were evolved with the \mbox{Collapse05v3} 
for the cases $Q/\sqrt{G} \mu=0,\,\,0.5,\,\,0.8$ (see  Sec. \ref{sec:tempevol}).   

The Fig. (\ref{fig:massadens}) shows the mass of the models as a function of its central density. 
We found that for any value of the charge below the extremal value ($Q/\sqrt{G} \mu<1$), there is still
unstable and  stable branches in the solutions.
 As it is well known from the first order perturbation theory the regions of the curve where $dM/d\rho <0$ represent  unstable solutions, and where
 $dM/d\rho \ge 0$ there are stable or marginally stable stars (see \cite{shapiro}, \cite{zeldovich}). 

There are models that have 
negative binding energy (see Fig. \ref{fig:massadens}b), but they are on the unstable branch of the solutions (see Fig. \ref{fig:massadens}a).

The maximum  mass models are indicated in the Fig. (\ref{fig:massadens}a) with a vertical bar, while the  tag ${\it 1st}$  or ${\it 2nd}$ show the position of the first or the second mass maximum, respectively.
The  second maximum in the curves, are due 
to a  pure relativistic effect. The reason is that
``energy has weight", paraphrasing Zeldovich and Novikov \cite{zeldovich}:  as the number of baryons increase
the total mass of the stars also increases attaining the first mass maximum. Passing the first maximum, the mass begins to decrease as the pressure became softer  at relativistic energies. However, with a further increase of the density the contribution to the ``weight'' due to the kinetic energy of the particles is more important respect to  the rest mass, and the second maximum appears. 

From the Fig. (\ref{fig:massadens}a), it is evident that this effect also takes place for charged neutron stars (see also tables \ref{tab:table1}-\ref{tab:table3}).

\begin{figure}[h]
\epsfig{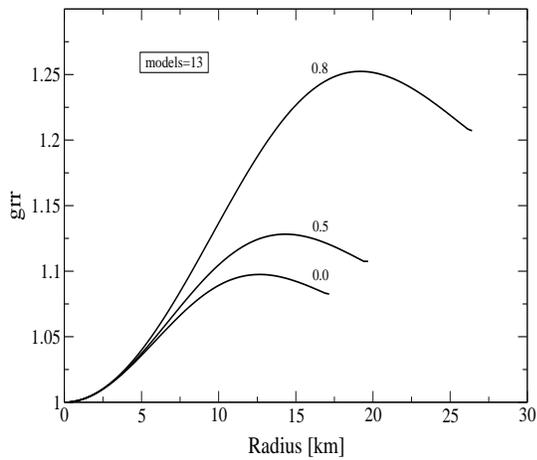}
\caption{\label{fig:metricam13} Coefficient $grr$ of the metric for models number $13$ 
and for
 three different
values of the charge to mass ratio $Q/\sqrt{G} \mu=0,\,\,0.5,\,\,0.8$.} 
\hspace{1cm}
\end{figure}

The Fig. (\ref{fig:metricam13}) shows 
the coefficient $grr$ of the metric as a function
of the lagrangian coordinate,
for model number $13$, for
 three different values of the charge to mass 
ratio $Q/\sqrt{G} \mu=0,\,\,0.5,\,\,0.8$.

In the Table \ref{tab:table1} are shown the results of the numeric integration of neutron star models without charge. 
The results are in agreement with the Oppenheimer and Volkov calculations. 
The first and second mass maximums are indicated on the table. 

In the tables \ref{tab:table2} and \ref{tab:table3}, are the results of the integration of neutron star models, with 
$Q/\sqrt{G}\mu=0.8$ and $Q/\sqrt{G}\mu=0.97$ respectively. 
The Tables \ref{tab:table2} and \ref{tab:table3} let us make a quantitative 
comparison of the charged neutron stars models
with the properties of the  neutron stars with zero charge, given in Table \ref{tab:table1}. 

In particular, the maximum mass for models with $Q/\sqrt{G}\mu=0.97$ is  $8.734\,M_{\odot}$, and the radius of this star is $75\,$km (see Table \ref{tab:table3}). 
For comparison, this case corresponds to  models with a parameter
$f=0.00111592$, in the notation of  Ray et al. \cite{ray}.
But they used a  charge distribution  proportional to the  mass-energy density and a different equation of state. However, 
the results presented in this section are in qualitative agreement 
with their results. 

We must observe that to study the possibility of 
making extremal Reissner-Nordstr\"om black holes it doesn't matter if the charge distribution is proportional to the rest mass density or total energy density, since the binding energy per nucleon tends to zero at the extremal case (see sub-sec. \ref{subsec:bindenerg}).

\subsubsection{Extremal case \label{extcase}}
With the HE05v1 it is possible to reach the sector $Q/\sqrt{G}\mu \ge 1$.
Black holes with this charge to mass ratio constitute naked singularities. 
 
We found that approaching the value $Q/\sqrt{G}\mu=1$, the mass and radius
of the models tends to infinity, within the computer capacity. 
Of course, it is impossible to show plots of this results, 
but a new technique was developed that will let us study this ``solutions" and will be presented in a separated paper \cite{ghezzi3}.

The  Newtonian Chandrasekhar's mass formula for charged stars \cite{ghezzi2} also predicts an infinite mass for the extremal case.
With a charge to mass ratio
 $\alpha=Q/\sqrt{G} \mu$, the
Chandrasekhar's mass is \cite{ghezzi2}:
\begin{equation}
M_{ch}=  5.83 \frac{Y_e^2}{(1-\alpha^2)} M_\odot\,.
\end{equation}
This equation reduces to the known Chandrasekhar's mass formula when $\alpha=0$ (no charge). 
For the extremal case $\alpha\rightarrow 1$, the formula gives a mass tending to infinite.

In the Sub-sec. \ref{sec:tempevol}, we will discuss the temporal evolution of neutron stars with extremal electric charge.

\subsubsection{\label{subsec:bindenerg}Binding energy per nucleon}
The  gravitational binding energy is given by
the difference between the total rest mass $\mu$ and the total gravitational mass $m$, \cite{wald}, \cite{glendenning}:
\begin{equation}
B=(\mu-m)\, c^2\,.
\end{equation}

The binding energy per nucleon is 
\begin{equation}
\frac{B}{A}=\frac{(\mu-m)\,  c^2}{A}
=\frac{m_n}{\mu}\,(\mu-m)\,  c^2\,,
\end{equation}
where $m_n$ is the rest mass of the nucleons, $c$ is the speed of light, and $A=\mu/m_n$ is the total number of nucleons on the star.

The total binding energy of the maximum mass model 
increase with the  charge 
(see Fig. \ref{fig:massadens}a). 

However, the contrary is true for the binding energy
per nucleon, i.e. it is lower with higher total charge.
For example, the values of $B/A$ for the maximum mass models are: 
$39.11$ Mev per nucleon for 
a neutron star with zero charge; $32.3$ Mev per nucleon
for a star with $Q/\sqrt{G} \mu=0.5$; $19.63$ Mev per nucleon for a star with $Q/\sqrt{G} \mu=0.8$;   
$7.46$ Mev per nucleon for a star with $Q/\sqrt{G} \mu=0.97$, and tending to  zero as $Q$ approaches the extremal value. 
For comparison, the binding energy per nucleon in the most stable finite atomic nucleus is roughly $\sim 8$ Mev per nucleon (see, for example, \cite{glendenning}).
It could be interesting to ask whether
a nearly extremal charged star, would disintegrate emitting high energy charged nucleus to strength its binding energy per nucleon. 
We will leave this question aside by now, since a more realistic EOS must be implemented.

\subsection{\label{sec:tempevol}Temporal evolution of charged neutron stars}
With the \mbox{Collapse05v3} it is possible to follow the temporal evolution of the stars.
 We arbitrarily choose to study the evolution of three models ($8, 13, 21$) constructed with the HE05v1 for each of the charge to mass values: $Q/\sqrt{G} \mu=0,\,\,0.5,\,\,0.8$.
Some of these models are on the stable branch, while others  are on the unstable branch (see Fig. \ref{fig:massadens}). The region passing the first maximum of the mass, where $dM/d\rho<0$, corresponds to
unstable models at first order in perturbation analysis. 
We do not study the evolution of models passing the second
mass maximum (see Fig. \ref{fig:massadens}, and tables
\ref{tab:table1}-\ref{tab:table3}).

The results of the simulations are summarized in the table
\ref{tab:table4}.

It was checked that the stars on the stable branch, say for example model $13$ with $Q/\sqrt{G}\mu=0.8$, could not jump to the unstable branch given a strong initial perturbation to the star.
The stable stars oscillate when perturbed and its (fundamental) 
frequency of oscillation was calculated 
(see table \ref{tab:table4}).

In the cases where a star oscillates we follow its 
evolution over
several periods of time to be sure that the equilibrium 
 is not a metastable state.
For a few models, we simulate the oscillating star over roughly 
one minute of physical time (that corresponds to several hours of computer time). Over this period of time,  the amplitude and the speed
of the oscillations are reduced due to the 
numerical viscosity. Of course,  the simulation can be extended so far as wanted in time. Eventually, the velocity will be  zero everywhere, and the star will rest in perfect hydrostatic equilibrium.  

The models number $8$ and $13$ are  stable  and they oscillate when perturbed.
We found that the frequency of oscillation is higher
for lower total charge (see table \ref{tab:table4}).  

Some of the stable charged stars collapses after a 
sudden discharge
of its electric field (see subsection below, and table \ref{tab:table4}). 

The charged stars
on the unstable branch
 collapses in agreement with the relativistic stellar perturbation theory (see for example \cite{shapiro}). 

It is possible 
to simulate the formation of an apparent horizon (and a trapped surface) 
and when this happens it is assumed that the star collapsed (see the section \ref{subsubsec:AH}).
In order to keep the accuracy of the results
the simulations are stopped when an apparent horizon forms (see the section \ref{sechydro}). 

As we will see in Sec. (\ref{subsubsec:AH}), 
 for each mass coordinate $\mu$ there is  a value of the coordinate $R$, denoted as $R_+$:
\begin{equation}
\label{rmas}
R_{+}(t,\mu)=\frac{G m(t,\mu)}{c^2} + \frac{G}{c^2} \sqrt{m^2(t,\mu)-\frac{Q^2(\mu)}{G}}\,,
\end{equation}
such that a trapped surface forms at coordinates $(t,\mu,\theta,\phi)$ iff  
\mbox{$R(t,\mu) \le R_+(t,\mu)$}  (see Sec. \ref{subsubsec:AH}).
In the Eq. (\ref{rmas}),  $m(t,\mu)$ and $Q(\mu)$ are the total mass and total charge at coordinate $\mu$, 
respectively, 
 $G$ is the gravitational constant 
and $c$ is the speed of light. 

If the perturbed energy is such that the total binding energy of an unstable star is greater than its equilibrium value,  the star will expand first and later collapse. On the contrary, if the  perturbed binding energy of the unstable star is lower than its equilibrium value, the star will collapse directly to a black hole.
The perturbed stable stars, will evolve through the path that carries it to the corresponding equilibrium point
on the equilibrium curve (see Fig. \ref{fig:massadens}). As the star
contracts or expands its central density changes accordingly, and  its evolutionary path  is  an horizontal segment in the graph of the total mass {\it versus} central density, or total binding energy {\it versus} central density (Figs. \ref{fig:massadens}). For the collapsing models this segment will extend to higher densities until the simulation is stopped.

All the models constructed with the \mbox{HE05v1} code and tested with the \mbox{Collapse05v3}
code have a charge $Q<\sqrt{G}\mu_s$, but for a few models we increase the charge
to the extremal value $Q=\sqrt{G}\mu_s$ (conserving the total energy). 
This is easily done on the \mbox{Collapse05v3}.
The result is that the star exploded, resulting in an  outward velocity at all the Lagrangian points (with the exception at the coordinate origin where the boundary condition is maintained: $u(\mu=0)=0$). This is another confirmation of the results obtained with the HE05v1 code (see subsection \ref{extcase}) that is not possible to get a finite mass star with extremal charge and bounded in a finite spatial region.

 The models number $21$ are in the unstable
branch (see Fig. \ref{fig:massadens}), and we found that all of them collapse (see table \ref{tab:table4}) independently of the amount of charge (with $Q < \sqrt{G} \mu_s$). However, the time at which the apparent horizon forms is higher for larger total charge in the star. 
 In all the cases studied there is no ejected matter.

\begin{figure*}[h]
\begin{center}
$\begin{array}{c@{\hspace{1in}}c}
\multicolumn{1}{l}{\mbox{\bf (a)}} &
\multicolumn{1}{l}{\mbox{\bf (b)}} \\ [-0.53cm]
\epsfig{width=7cm,height=6cm, file=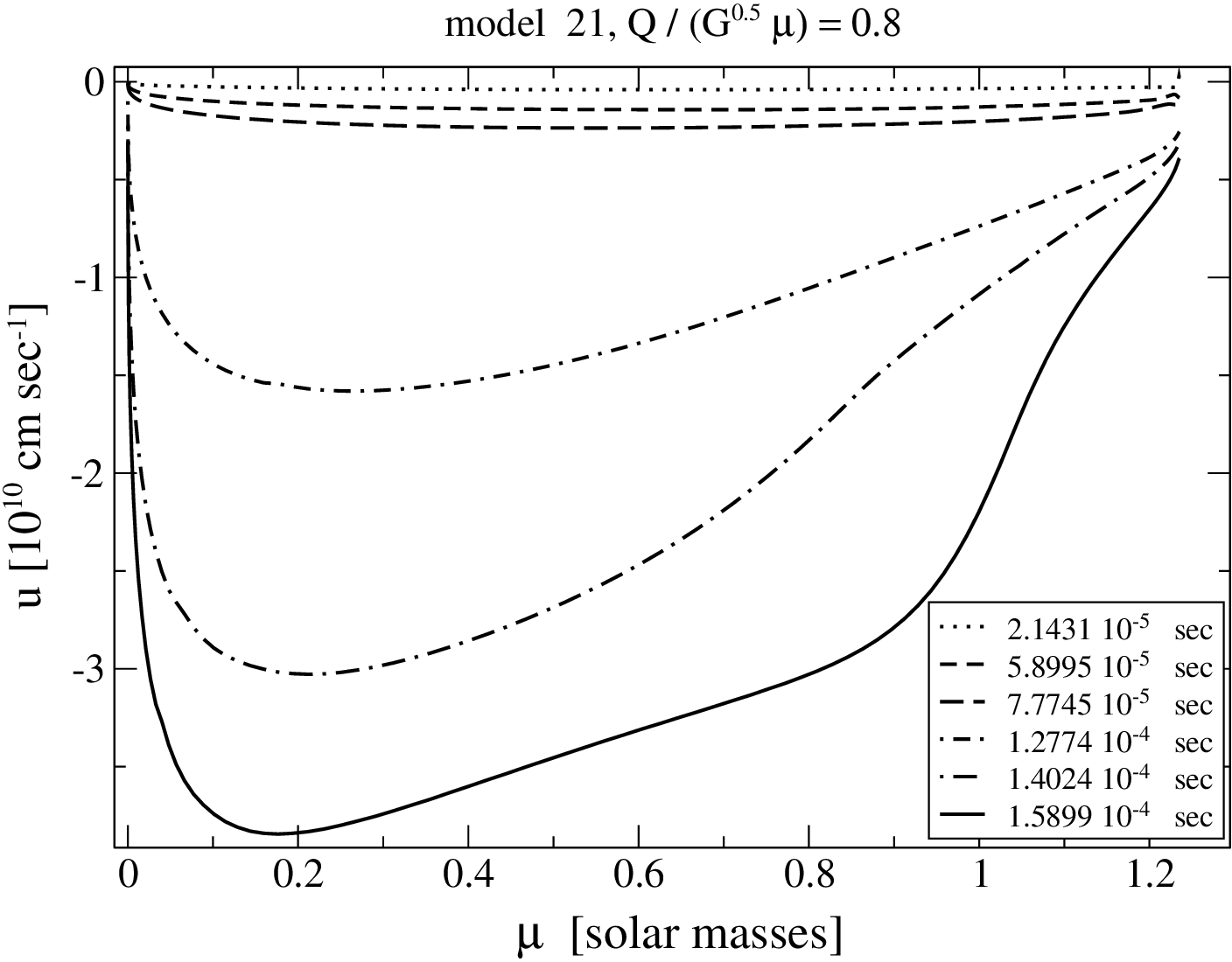} &
\epsfig{width=7cm,height=6cm, file=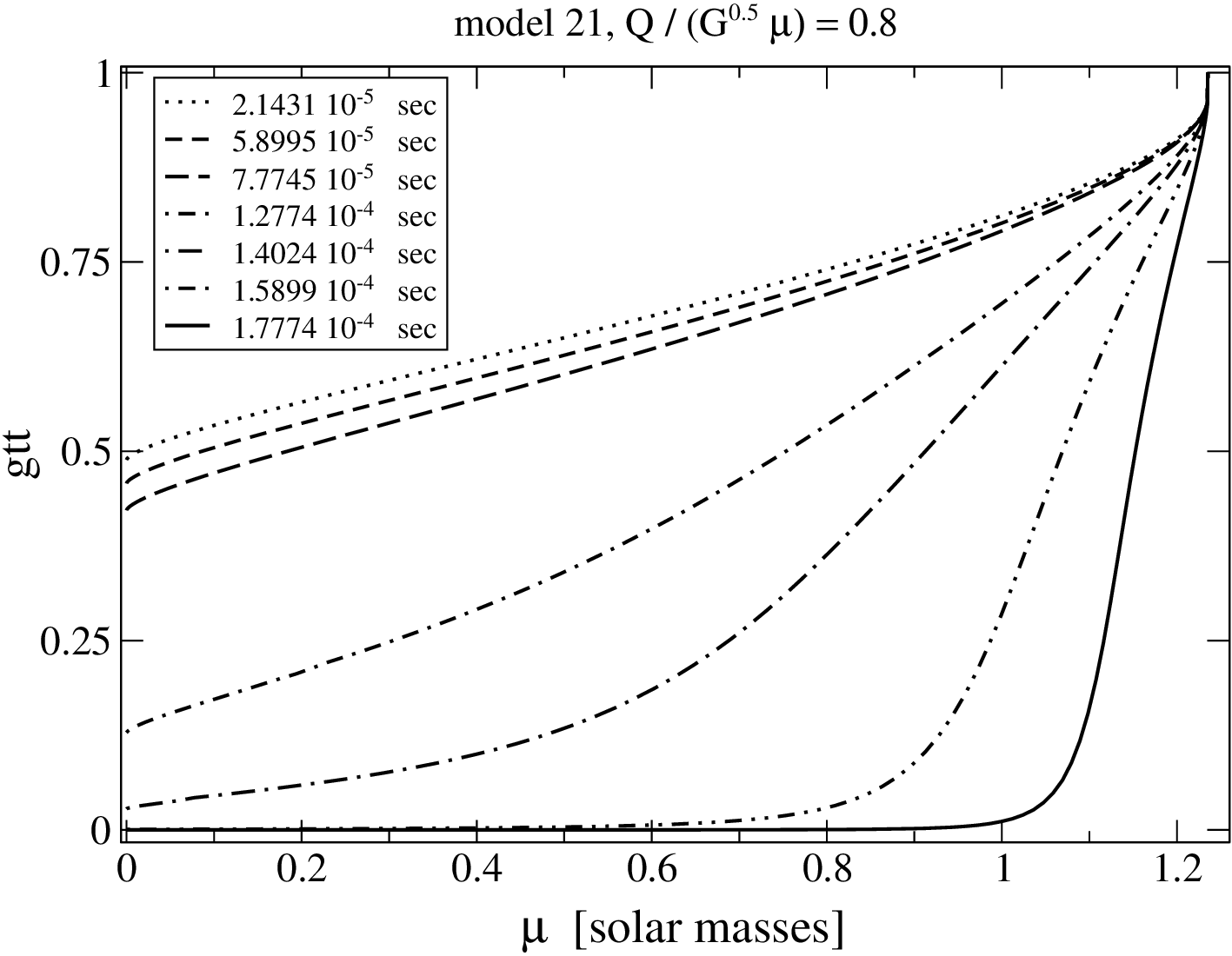}
\\ [0.4cm]
\mbox{\bf (a)} & \mbox{\bf (b)}
\end{array}$
\end{center}
\caption{Temporal snapshots of the speed $u$, for different time-steps (Fig. a), and snapshots of the coefficient $g_{tt}$ of the metric, for different time-steps (Fig. b), for   
model 21 and for a charge to mass ratio $Q/\sqrt{G} \mu=0.8$.}
\label{fig:v08m21}
\end{figure*}

In the figures it is shown the effect of these perturbations by means of a series of temporal snapshots. 
The  Fig. (\ref{fig:v08m21}) corresponds to the collapsing model $21$ with $Q/\sqrt{G}\mu=0.5$.
In the Fig. (\ref{fig:v08m21}a) it is plotted a series of temporal snapshots
of the velocity profiles for this model.
 We see that the in-falling matter acquires a relativistic speed and 
there are not strong shock waves formed.
The  Fig. (\ref{fig:v08m21}b) shows a series of temporal snapshots
of the metric coefficient $g_{tt}$ as function of the coordinate $\mu$.
The metric
coefficient $g_{tt}$ goes to zero in this case, 
which is a sufficient condition for the formation of an apparent horizon 
(see Sec. \ref{subsubsec:AH}).
The Fig. (\ref{fig:v05m21}) corresponds to the model $21$ with $Q/\sqrt{G}\mu=0.5$.
This is another example of an star that collapses without forming strong shock waves
(see  Fig. \ref{fig:v05m21}a). The results shown in the Fig. (\ref{fig:v05m21}b) also 
assures that the star collapsed to a charged black hole (see Sec. \ref{subsubsec:AH}).

\begin{figure*}[h]
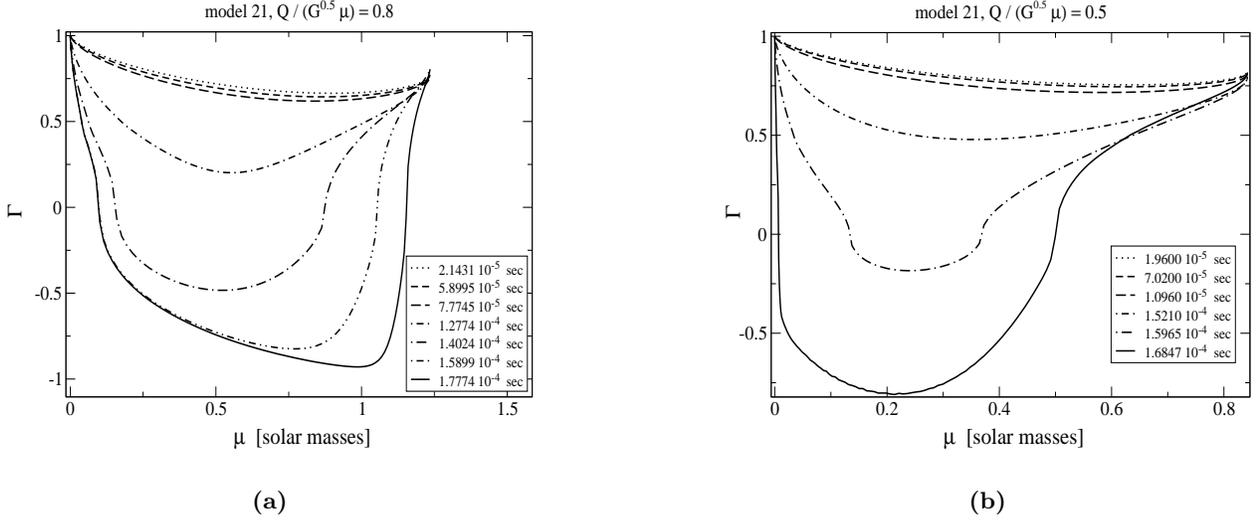

\begin{center}
$\begin{array}{c@{\hspace{1in}}c}
\multicolumn{1}{l}{\mbox{\bf (a)}} &
\multicolumn{1}{l}{\mbox{\bf (b)}} \\ [-0.53cm]
\epsfig{width=7cm,height=6cm, file=PRDIIgamma08m21.eps} &
\epsfig{width=7cm,height=6cm, file=PRDIIgamma05m21b.eps}
\\ [0.4cm]
\mbox{\bf (a)} & \mbox{\bf (b)}
\end{array}$
\end{center}
\caption{Temporal snapshots of the factor $\Gamma$, for different time-steps, for
model 21 and for a charge to mass ratio $Q/\sqrt{G} \mu=0.8$ (Fig. a), and temporal snapshots of the factor $\Gamma$,
 for different time-steps, for
model 21 and for a charge to mass ratio $Q/\sqrt{G} \mu=0.5$ (Fig. b). }
\label{fig:gammas}
\hspace{1cm}
\end{figure*}

The Fig. (\ref{fig:gammas}) shows a comparison of the evolution of 
the factor $\Gamma$ for  two simulations of the model $21$
 with charge to mass ratios $Q/\sqrt{G}\mu=0.8$ (Fig. \ref{fig:gammas}a), 
and $Q/\sqrt{G}\mu=0.5$ (Fig. \ref{fig:gammas}b). 
When the function $\Gamma$ approachs zero
the fluid enters a regime called of ``continued collapse'', in this case 
the gradient of pressure and of electric charge is no more
effective in counterbalancing the gravitational attraction,
the fluid is almost in free fall and nothing
can stop the collapse. 
From the Eq. (\ref{dynam}) we see that the first term inside the
brackets is nearly zero in this regime. Moreover,
  the figures (\ref{fig:gammas}a) and  (\ref{fig:gammas}b) 
indicate that the factor $\Gamma$ can acquire
negative values after the formation of an apparent horizon,
in this case the Eq. (\ref{dynam}) indicate that this behaviour
reinforce the collapse.

\begin{figure*}[h]
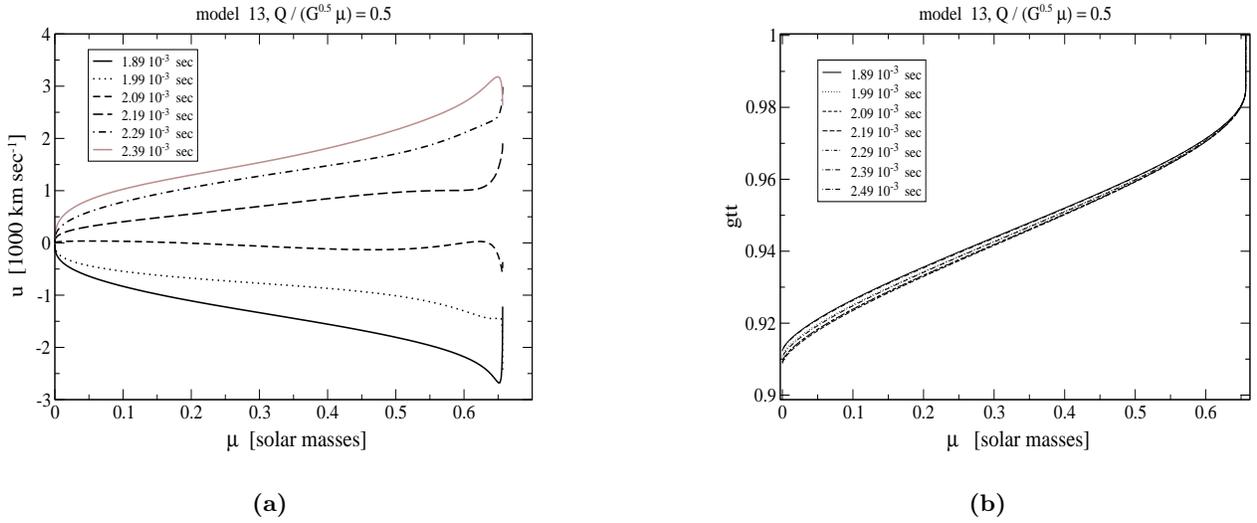

\begin{center}
$\begin{array}{c@{\hspace{1in}}c}
\multicolumn{1}{l}{\mbox{\bf (a)}} &
\multicolumn{1}{l}{\mbox{\bf (b)}} \\ [-0.53cm]
\epsfig{width=7cm,height=6cm, file=PRDIIvel05m13.eps} &
\epsfig{width=7cm,height=6cm, file=PRDIIgtt05m13.eps}
\\ [0.4cm]
\mbox{\bf (a)} & \mbox{\bf (b)}
\end{array}$
\end{center}
\caption{Snapshots of the speed $u$ (Fig. a) and snapshots of the coefficient $g_{tt}$ of the metric (Fig. b), for different time-steps over half period of oscillation, for
model 13 and for a charge to mass ratio $Q/\sqrt{G} \mu=0.5$.}
\label{fig:v05m13}
\end{figure*}

\begin{figure*}[h]
\begin{center}
\epsfig{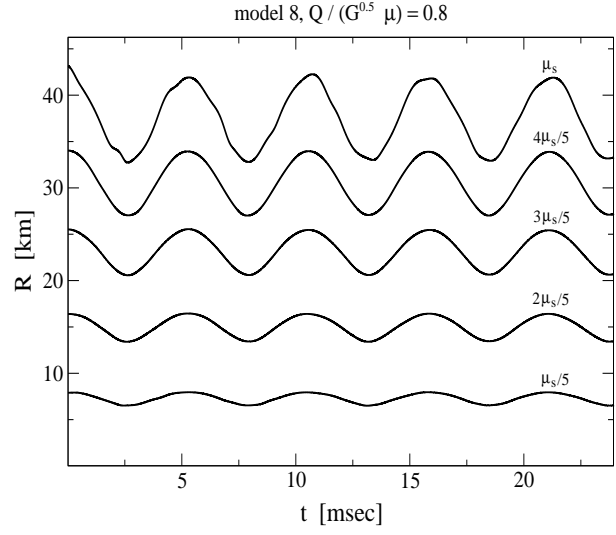} 
\end{center}
\caption{Temporal evolution of several layers of an oscillating star, 
for model $8$ and with charge to mass ratio $Q/\sqrt{G} \mu=0.8$, over several oscillation periods. 
Each curve is labeled by the quantity of rest mass enclosed, or equivalently by its
Lagrangian coordinate $\mu$, given in units of the total rest mass $\mu_s$}
\label{fig:oscil}
\end{figure*}

The Fig. (\ref{fig:v05m13}) corresponds to the oscillating model $13$ with $Q/\sqrt{G}\mu=0.5$.
 In this
case the amplitude of the velocity profiles are bounded, roughly between the values
$\pm 3000\,\,{\rm km\,\,sec}^{-1}$  (Fig. \ref{fig:v05m13}a).
The factor 
$g_{tt}$ shows very little
variation in this case (Fig. \ref{fig:v05m13}b). 
The Fig. (\ref{fig:oscil}), shows the temporal evolution of five layers of the
star, which oscillate over several periods. The mass enclosed by the layers is indicated in the figure.

\begin{table*}
\caption{\label{tab:table4} Charged neutron star models evolved forward in time.}

\begin{ruledtabular}
\begin{tabular}{rccccccc}

Model\footnotemark[1]& $Q/\sqrt{G}\mu$ & Total Rest Mass $\mu$ & Binding Energy & Oscillation frequency\footnotemark[2] & Collapse time\footnotemark[3] &fate\\
& & $M_{\odot}$ & $M_{\odot}\,c^2$ & Hz  & Sec & &\\
\hline
21 &    0.0 &  0.679 & 0.021 & - & $1.6152 \times10^{-4}$ &  collapse&\\
21 &    0.5 &  0.825 & 0.015 & - & $1.6847 \times10^{-4}$ &  collapse&\\
21 &    0.8 & 1.237 & $-0.008\hspace{2.48mm}$ & - & $1.7774 \times10^{-4}$ & collapse&\\
13 &    0.0 &  0.449  & 0.009 & $1333\hspace{1.6mm}$ & - & oscillate&\\
13 &    0.5 &  0.655  & 0.013 & 667 & - & oscillate&\\
$13'\hspace{-0.9mm}$ &    0.5 &  -  & - & $1000\hspace{1.6mm}$ & - & oscillate&\\
13 &    0.8 &  1.600   & 0.027 & 490 & - & oscillate&\\
$13'\hspace{-0.9mm}$ &    0.8 &  -   & - & - & $2.3595 \times10^{-4}$ & collapse&\\
8 &    0.0 &  0.169   & 0.001 & 476 & - & oscillate&\\
8 &    0.5 &  0.258   & 0.002 & 249 & - & oscillate&\\
$8'\hspace{-0.95mm}$ &    0.5 &  -   & - & 470 & - & oscillate&\\
8 &    0.8 &  0.749   & 0.007 & 200 & - & oscillate&\\
$8'\hspace{-0.95mm}$ &    0.8 &  -   & - 
& - & $9.7525 \times10^{-4}$ & collapse&\\
- &    $\ge 1.0\hspace{0.33cm}$ &  -   & - & - & - & explodes\footnotemark[4]$\hspace{-0.17cm}$&\\
 
\end{tabular}
\end{ruledtabular}
\footnotetext[1] {The primed numbers indicate models that
were suddenly discharged in the simulation.}
\footnotetext[2]{The oscillation frequency of the stable models depends very weakly  on the numerical viscosity.
When the numerical viscosity coefficient is divided by a factor of two, there is no appreciable change in the frequency.}
\footnotetext[3]{The collapse time, is the time elapsed between the formation of the apparent horizon and the beginning of the simulation.}
\footnotetext[4]{Any model with $Q/\sqrt{G}\mu \ge 1.0$
explodes, i.e., the matter scatters to infinity.}
\end{table*}

\subsubsection{\label{subsubsec:AH} The formation of an apparent horizon}
A trapped surface is a space-like orientable two dimensional compact surface,
such that inward and outward null geodesics normal to it
converge. The apparent horizon is the outer boundary of all the trapped surfaces (see \cite{penrose}, \cite{mashhoon}, \cite{poisson2}, \cite{wald}, \cite{novikov},  and \cite{hawking}, for definitions and related
theorems).  
A black hole in asymptotically flat space-time is the region from where no causal or 
light signals can reach  $\mathcal{I}^{+}$ (the future null infinity, see for 
example \cite{wald} and \cite{novikov} for definitions). The event horizon $H^{+}$ 
is the boundary of the black hole region $H^{+}=\dot{J}^{-}(\mathcal{I}^{+})$.

  The formation of an apparent horizon is a sufficient, although not necessary, 
condition for the formation of a black hole (see Ref. \cite{novikov}).  
In fact, in the case an apparent horizon forms  the theorems of Hawking 
and Penrose (1970) can be applied to know that the outcome will be the
formation of a singularity, i.e.: the spacetime contains 
at least one incomplete timelike or null geodesics (see the theorems 9.5.3 and 9.5.4 in 
\cite{wald}, Pags. 239-241, known as ``singularity theorems''). 
Moreover,  the trapped surface (and the apparent horizon) 
is contained within a black hole as
a mathematical proposition asserts (see the propositions 12.2.2 and 12.2.3 in \cite{wald}, 
Pags. 309-310, and 
proposition 9.2.1 in \cite{hawking}, Pag. 311). 

So, we assumed that if an apparent horizon is formed in the stellar collapse, the result will be the inexorable formation of a black hole.

In a numeric simulation, in general, is needed an algorithm to find out the apparent horizon's location. However, in the present study it is easy to guess that the trapped surfaces are spheres. 
With this in mind, we can describe the apparent horizon formation and evolution. 
The vector  $l^{\nu}=((a c)^{-1},b^{-1},0,0)$ is tangent to the outgoing null geodesics, 
and   $n^{\nu}=((a\, c)^{-1},-b^{-1},0,0)$ is tangent to the ingoing null geodesics, in co-moving coordinates. They are also orthogonal to the constant $(t,\mu)$ surfaces.  
Following Mashhoon and Partovi \cite{mashhoon}, we define the quantities $\Psi=l^{\nu} R_{,\nu}$ and $\Phi=n^{\nu} R_{,\nu}$ which are related by a positive multiplicative factor to the expansion factor
for radially ingoing and outgoing null geodesics, respectively. 
Thus, the  coordinates of the apparent horizon are found solving the equation: 
\begin{equation}
\label{apparenteq}
\Psi(t,\mu)=(a\,c)^{-1} R_{,t} + b^{-1} R_{,\mu}=0\,,
\end{equation} 
 using the Eqs. (\ref{gammadef}) and (\ref{udefin}), and squaring, this is equivalent to 
\begin{equation}
\Gamma^2=\frac{u^2}{c^2}\,,
\end{equation} 
and using Eq. (\ref{gamma2}), this gives:
\begin{equation}
1-\frac{2 m G}{R c^2}+
\frac{G Q^2}{c^4 R^2}=0\,.
\end{equation} 
The solution of this equation is:
\begin{equation}
\label{ahrp}
R_{\pm}(t,\mu)=\frac{G \,m(t,\mu)}{c^2} \pm \frac{G}{c^2} \sqrt{m^2(t,\mu)-\frac{Q^2(\mu)}{G}}\,,
\end{equation}
where $R_{-}$ is the inner boundary, while $R_{+}$ is the outer boundary of the trapped surfaces. So, the apparent horizon is the 
surface (in Schwarszchild coordinates):
 $$S_{AH} : R=R_{+},\,\,\theta=\theta,\,\,\phi=\phi\,.$$
In order to know the space-time coordinates of the apparent horizon in co-moving
coordinates, we must solve the Eq. (\ref{ahrp}) for $t$ and $\mu$. This is done 
with a numeric subroutine that for each time $t$ and for each coordinate $\mu$  asks
if $R(t,\mu) = R_+(t,\mu)$ \footnote{The apparent horizon is not stationary, it evolves with time. The equation of motion for the apparent horizon was found by Bekenstein \cite{bekenstein} and by Mashhoon and Partovi \cite{mashhoon}.}.

Some of the stars when evolved in time give rise to the formation of an apparent horizon, that is the coordinate $R$ contracted to
the value $R_{+}$ for some $t$ and $\mu$.
The table \ref{tab:table4} summarize the results of the simulations: it is shown that some of the stars collapsed, and it is given the time elapsed from the beginning of the simulation and the formation of the apparent horizon.

The Eq. (\ref{apparenteq}) is also equivalent to:
\begin{equation}
\frac{d\mu}{dt}=\frac{a}{b}\,c\,,
\end{equation}
this equation can also be obtained from the equation for light rays $ds^2=0$ \footnote{For ingoing light rays the equation has a minus sign. See also Eq. (33) in ref. \cite{may}}.
From this equation we see that outgoing light rays have zero expansion in the co-moving frame 
when $a=g_{tt} \rightarrow 0$ \footnote{This is a sufficient but not necessary condition 
on the formation of the apparent horizon}. However, from the simulation was 
obtained that the apparent horizon forms before $a=0$, this means 
that radial light rays enter the trapped region (at its boundary $S_{AH}$) with $d\mu/dt\ne0$.  

In the Figs. (\ref{fig:v08m21}b) and (\ref{fig:v05m21}b) is shown the evolution of the metric coefficient $a$ for  model  number $21$ with $Q/\sqrt{G}\mu=0.5$, and for model
number $21$ with $Q/\sqrt{G}\mu=0.8$ (this model  is
one of the entries of the table \ref{tab:table2}). As we see,  $a \rightarrow 0$ for this models, so they collapse. 

The event horizon can not be described in the simulations, since it is the result of the complete history of the collapsing matter (while the simulation lasts a finite amount of time). But we know that the trapped region is contained in the black hole region. There exists only one
situation in which the formation of
the event horizon can occur in a finite coordinate time: when all the matter  composing the star collapses in a finite time.
Only in the special case that the star collapse completely, i.e.: if $R(t_0,\mu_s) \le R_{+}(t_0,\mu_s)$ for some finite time $t_0$,  the apparent horizon will coincide with the event horizon of a Reissner-Nordstr\"om space-time:
$R_{BH}(M,Q) = R_{+}(t_0,\mu_s)$, where $R_{BH}$ is the Reissner-Nordstr\"om radius of a black hole of total mass-energy $M$ and total charge $Q$. 
In all the simulations performed the apparent horizon is formed for some coordinate $\mu$ with $0<\mu<\mu_s$. So, 
in the present set of simulations, the apparent horizon never coincides with the event horizon. 

\subsubsection{Evolution of suddenly discharged stars}
We picked up some stable stars and followed their temporal evolution
after a sudden discharge of its electric field. The scenario is a pure hypothetical  one, in which the neutron star is leaved in a charged  metastable state after its formation, or during an accretion process onto it. The charged matter could suffer a discharge 
by recombination of charges of opposite sign, or by effect of the high electrical conductivity.
We emphasize that  we are not claiming that this is a possible astrophysical scenario, but we want to explore the different theoretical possibilities for the stellar dynamics.

The simulation is performed with the worst conditions for the stellar stability: the electric field is simply turned off after the first time-steps of temporal evolution of an otherwise stable  star. The energy is maintained constant in the process, corresponding to a conversion of the electromagnetic energy into heat or internal energy.
This is easily performed with the \mbox{Collapse05v3} code.

The results of the simulation of the stars suddenly discharged are summarized in the table \ref{tab:table4}. The models discharged are 
indicated with primed
numbers. We see that not all the models collapse to form black holes: the model number $13$ with an initial charge to mass ratio $Q/\sqrt{G} \mu =0.5$ oscillate, while the same model with $Q/\sqrt{G} \mu =0.8$ collapsed
after a sudden discharge; the model number $8$ oscillate if its initial charge to mass ratio is $Q/\sqrt{G} \mu =0.5$, while the same model collapses if $Q/\sqrt{G} \mu =0.8$. 
The models number $21$ always collapse, as they fall on the unstable branch, then it is unnecessary to consider them.

\subsubsection{The formation of a shell}
In the Reference \cite{ghezzi2} it is described  the formation of a shell of higher density formed near the surface of the star. Although this effect must happen in Newtonian physics, its evolution in the strong field regime is highly non-linear and far from obvious.

The weight of the star is supported
by the gas pressure and by the Coulomb repulsion of the
matter. If the Coulomb repulsion is important respect to the
gravitational attraction -although not necessarily stronger- a shell of matter can form (see 
\cite{ghezzi2} for an explanation).
The contrast of density between the shell and its interior depends on the
energy and charge distribution.
So, it is important to check if the effect take place in the present simulations.
 
We found that the shell
forms only mildly, or do not form, and its late behavior can not be followed clearly.

From all the numerical experiments  performed, it is observed 
that the shell formation effect arise more clearly when the initial
density profile is flat (like the simulations of Ref. \cite{ghezzi2}).
In consequence, the shell formation is not an important physical
effect in the collapse of charged neutron stars. However, it could be
important for the collapse of the core of super-massive stars \cite{ghezzi2}.

\vspace{.45cm}

\begin{figure*}[h]
\begin{center}
$\begin{array}{c@{\hspace{1in}}c}
\multicolumn{1}{l}{\mbox{\bf (a)}} &
\multicolumn{1}{l}{\mbox{\bf (b)}} \\ [-0.53cm]
\epsfig{width=7cm,height=6cm, file=PRDIIvel05m21b.eps} &
\epsfig{width=7cm,height=6cm, file=PRDIIgtt05m21.eps}
\\ [0.4cm]
\mbox{\bf (a)} & \mbox{\bf (b)}
\end{array}$
\end{center}
\caption{Temporal snapshots of the speed $u$, for different time-steps (Fig.a) and snapshots of the coefficient $g_{tt}$ 
of the metric, for different time-steps (Fig. b), for   
model 21 and for a charge to mass ratio $Q/\sqrt{G} \mu=0.5$.}
\label{fig:v05m21}
\hspace{1cm}
\end{figure*}

\section{Final Remarks\label{secremarks}}
It is usually assumed in astrophysics that stars haven't important internal
electric fields. 
 Whether an star can have large internal electric fields or a net total charge is not yet clear.
However, several features could be common to the evolution of rotating collapsing stars, 
with the angular momentum playing the role of the electric charge. 
So, the  present study can shed some light on more realistic astrophysical scenarios. 
 
In the models studied in this paper we assumed, for simplicity, that the charge density is
proportional to the rest mass density.
We found that in hydrostatic equilibrium
the charged  stars have a larger mass and radius
than the uncharged ones. This is, as expected, due to the Coulomb repulsion.
 The mass of the models tends
to infinity as the charge approaches the extremal value
$Q=\sqrt{G} \mu_s$.  The hydrostatic equilibrium solutions with $Q\ge\sqrt{G} \mu_s$ gives models with an infinite mass and radius (see also \cite{ghezzi3}). This means that in this particular cases the integrated  mass and radius 
diverges within the computer capacity. 

All the models with charge less than the extremal  ($Q<\sqrt{G}\mu_s$), have a mass 
limit and there are unstable and stable solutions. We checked the stability
of the solutions integrating forward in time the models, and
applying  strong perturbations to them. Some of the models collapse directly to form black holes,  without ejecting matter. Other models oscillate. For a given model, with fixed central density, the frequency of oscillation is lower when the charge is higher. The frequency of oscillation is
weakly dependent on the numerical viscosity. The models that collapse are solutions with $dM/d\rho<1$, while the oscillating models have $dM/d\rho>1$. So, the stability of the models agree with the predictions of the first order perturbation theory \cite{shapiro}.

It seems that there is a limit for the charge that a star can have, which is lower than the extremal case. This limit
 arise because the binding energy per nucleon of the models with 
$Q/\sqrt{G}\mu\ge0.97$, is lower than the binding energy per nucleon on an atomic  nucleus. Then, it is possible that this stars disintegrate
to reach a more bounded energy state. This point deserves  further study.

From a pure theoretical point of view, the issue of the collapse of charged fluid spheres, 
is related to the third law of black hole thermodynamics \cite{israel}, and with the cosmic censorship hypothesis \cite{wald},
\cite{joshi}. 
The third law of black hole physics states that the temperature 
of a black hole cannot be reduced to zero by a finite number of operations.
The impossibility of transforming a black hole into an extremal one,
in a finite number of steps, is related to the impossibility of
getting $Q=\sqrt{G} \mu_s$ in some particular experiment \cite{israel}, \cite{novikov}. In agreement with this law, we found that 
it is not possible to form extremal black holes from the collapse
of a charged fluid ball.
In fact, any charged ball with a charge to mass ratio greater or equal than one explodes, or its matter spreads. 

The black holes with $Q\ge\sqrt{G} \mu_s$ represent naked singularities, and the
impossibility of getting black holes with this charge to mass ratio in the present simulations  are in agreement with the ``cosmic censorship hypothesis".

It must be remarked that once an apparent horizon forms the formation of an event horizon is inevitable, and as we
proved the matching of the interior solution with an exterior Reissner-Nordstr\"om solution, we simulated here
the formation of a Reissner-Nordstr\"om space-time from a gravitational stellar collapse.

Although we are not using a realistic equation of state
we think that the results are of general validity, at least
from a qualitative point of view. 

Other fields or more exotic physics must be considered in order to form extremal black holes.

\section{\label{App:appendix1}Appendix A}
For completeness we reproduce here the Einstein 
equations in co-moving coordinates as derived by
Landau and Lifshitz (see \cite{landau}, Pag. 311).
The equations of section \ref{sec2} were derived 
from these by making the replacements $a^2 c^2=e^{\nu} c^2$, $b^2=e^{\lambda}$, and
$R^2=e^{\mu}$.
The Einstein equations are:
\begin{eqnarray*}
&&\frac{8 \pi G}{c^4} T^1_1=\frac{8 \pi G}{c^4} \biggl(-P+\frac{Q^2}{4 \pi R^4}\biggr)= 
\frac{1}{2} e^{-\lambda} \biggl(\frac{\mu'^2}{2}+\mu' \nu' \biggr)\\ 
&&\hspace{1.5cm}-e^{-\nu} \biggl(\ddot{\mu} -
\frac{1}{2} \dot{\mu} \dot{\nu}+\frac{3}{4} \dot{\mu}^2 \biggr)-
e^{-\mu}\,,\\
&&\frac{8 \pi G}{c^4} T^0_0=\frac{8 \pi G}{c^4} \biggl(\delta c^2+\frac{Q^2}{4 \pi R^4}\biggr)= 
-\frac{1}{2} e^{-\nu} \biggl(\frac{\dot{\mu}^2}{2}+\dot{\mu} \dot{\lambda} \biggr) \\ 
&&\hspace{1.5cm}+e^{-\lambda} \biggl(\mu'' -
\frac{1}{2} \mu' \lambda'+\frac{3}{4} \mu'^2 \biggr) -e^{-\mu}\,,\\
&&\frac{8 \pi G}{c^4} T^1_0=0=\frac{1}{2}\,e^{-\lambda}\,
\biggl(-2 \dot{\mu}'-\dot{\mu} \mu'+\dot{\lambda} \mu'+\nu' \dot{\mu} \biggr)
\end{eqnarray*}

\section*{Acknowledgments}
The author thank Professor Dr. Jacob D. \mbox{Bekenstein} for 
clarifications about his equations.
This work was performed with the computational resources
of Professor Dr. Patricio S. \mbox{Letelier}, so the author thank him.
The author thank an anonymous referee for useful comments that help
to improve this paper.
The author acknowledge the Brazilian agency FAPESP
for support.

\end{document}